\documentclass[12pt,titlepage]{article}
\usepackage{geometry}                % See geometry.pdf to learn the layout options. There are lots.
\usepackage{graphicx}
\usepackage{amsmath, amssymb,wasysym} 
\usepackage{hyperref}
\usepackage[T1]{fontenc}
\usepackage{helvet}
 
\usepackage[round,numbers,sort&compress]{natbib}

\usepackage{geometry}
\usepackage{fleqn}
\usepackage{color}
\usepackage{setspace}

\newcommand{\avg}[1]{\langle{#1}\rangle}       
\definecolor{mygray}{gray}{.75}
\definecolor{mygreen}{rgb}{0.0,0.48,0.16}

\renewcommand{\figurename}{FIGURE}

\begin{document}

\title{Crowding of molecular motors determines microtubule depolymerization}

\author{Louis~Reese\thanks{Email:~louis.reese@physik.lmu.de} \and Anna~Melbinger\thanks{Email:~anna.melbinger@physik.lmu.de} \and Erwin~Frey\thanks{Corresponding author. Email:~frey@lmu.de} \\
\\ Arnold Sommerfeld Center for Theoretical Physics and \\ Center for NanoScience, Department of Physics, \\ Ludwig-Maximilians-Universit\"{a}t M\"{u}nchen, \\Theresienstra\ss e 37, D-80333 Munich, Germany}

\date{}

\date{\today}
\maketitle
\abstract{Assembly and disassembly dynamics of microtubules (MTs) is tightly controlled by MT associated proteins. Here, we investigate how plus-end-directed depolymerases of the kinesin-8 family regulate MT depolymerization dynamics. Employing an individual-based model, we reproduce experimental findings. Moreover, crowding is identified as the key regulatory mechanism of depolymerization dynamics. Our analysis gives two qualitatively distinct regimes. For motor densities above a particular threshold, a macroscopic traffic jam emerges at the plus-end and the MT dynamics become independent of the motor concentration. Below this threshold, microscopic traffic jams at the tip arise which cancel out the effect of the depolymerization kinetics such that the depolymerization speed is solely determined by the motor density. Because this density changes over the MT length, length-dependent regulation is possible. Remarkably, motor cooperativity does not affect the depolymerization speed but only the end-residence time of depolymerases.}

\vspace{1cm}

\section*{\bf INTRODUCTION}

Microtubules (MTs) are cytoskeletal filaments that serve a central role in intracellular organization \citep{HAY01,TOL10} and several cellular processes including mitosis~\citep{SHA00,KAR01}, cytokinesis \citep{EGG06} and intracellular transport \citep{HIR09}. They can cope with this multitude of diverse tasks because they are highly dynamic structures which continually assemble and disassemble through the addition and removal of tubulin heterodimers at their ends. GTP-hydrolysis is the energy source which drives switching between persistent states of growth and shrinkage, a stochastic process termed dynamic instability~\citep{MIT84,DOG93,DES97,HOW03}. Each cellular process employs a specific set of MT-associated proteins (MAPs) to tightly regulate the rates of growth and shrinkage as well as the rate of transition between these states \citep{WOR05,HOW07, HOW09}. 

Depolymerases from the kinesin-8 and kinesin-13 protein families (e.g., Kip3p and MCAK, respectively) are important regulators of MT dynamics. They are thought to promote switching of MTs from growth to shrinkage (catastrophes) \citep{HOW07}. Whereas MCAK lacks directed motility and diffuses along MTs \citep{HEL06}, Kip3p is a highly processive plus-end-directed motor \citep{VAR06,GUP06}. Proteins from the kinesin-8 family are important for regulating MT dynamics in diverse organisms. Kif18A is a key component in chromosome positioning in mammalian cells \citep{MAY07,STU08,DU10} where it regulates plus-end dynamics. Its orthologs, the plus-end directed motors Kip3p in budding yeast \citep{GUP06} and Klp5/6 in fission yeast \citep{UNS08,TIS09,GRI09}, show depolymerizing activity. A notable feature shared by these MT plus-end depolymerases is that they depolymerize longer MTs more rapidly than they do shorter ones~\citep{VAR06,MAY07,TIS09,VAR09}. A similar length-dependent regulation of MT assembly by kinesin-5 motors was observed in \emph{in vivo} studies of chromosome congression in budding yeast~\citep{GAR08}. The key experimental observations from \emph{in vitro} studies of Kip3p~\citep{VAR09} are that 1), the end-residence time of Kip3p at the tip depends on the bulk concentration of Kip3p and correlates inversely with the macroscopic depolymerization speed; and 2), the macroscopic depolymerization rate is directly proportional to the flux of Kip3p towards the MT plus-end. 

It is thought that length-dependent depolymerization kinetics serves several purposes~\citep{TOL10}. For example, positioning of the nucleus at the cell center during interphase is achieved by growing MTs that push against the cell poles while remaining attached to the nucleus. A higher rate of catastrophes for longer MTs implies that shorter MTs have an increased contact time with the cell poles. Computer simulations show that this leads to a higher efficiency of nuclear positioning during interphase \citep{FOE09}.

There is convincing experimental evidence that molecular traffic along MTs strongly affects MT depolymerization dynamics. However, \emph{in vitro} experiments can not yet fully explore the underlying traffic dynamics. Theoretical investigations employing individual-based models can be instrumental in furthering a mechanistic understanding of this process. Fortunately, these models can be constructed on the basis of substantial quantitative data available from \emph{in vitro} experiments \citep{VAR06,VAR09} characterizing the binding kinetics and the motor activity of plus-end-directed motors. Therefore we sought to identify the molecular mechanisms underlying the observed correlation between depolymerization dynamics and molecular traffic along MTs.

In this study, we constructed an individual-based model for the coupled dynamics of MT depolymerization and molecular traffic of plus-end-directed motors. This model quantitatively reproduces previous experimental results~\citep{VAR06,VAR09}. Moreover, we make precise quantitative predictions for the density profiles of molecular motors on the MT and demonstrate that molecular crowding and ensuing traffic jams regulate the depolymerization dynamics. We find two qualitatively distinct regimes of depolymerization dynamics: At low bulk concentrations of depolymerases, the depolymerization speed of MTs is density-limited and is a function of the bulk concentration and average motor speed alone. There is a sharp threshold in bulk depolymerase concentration above which macroscopic traffic jams emerge and the depolymerization speed is simply given by the microscopic depolymerization rate. Of note, none of these features are affected by the degree of cooperativity in the depolymerization kinetics. In contrast, the end-residence time of a depolymerase (i.e., the typical time it spends at the plus-end) is strongly correlated with cooperativity. We outline how these predictions from our theoretical analysis can be tested experimentally. 

\section*{\bf RESULTS}

\subsection*{\bf Model definition}

\begin{figure}[t]
\centering 
\includegraphics*[width=3.25in]{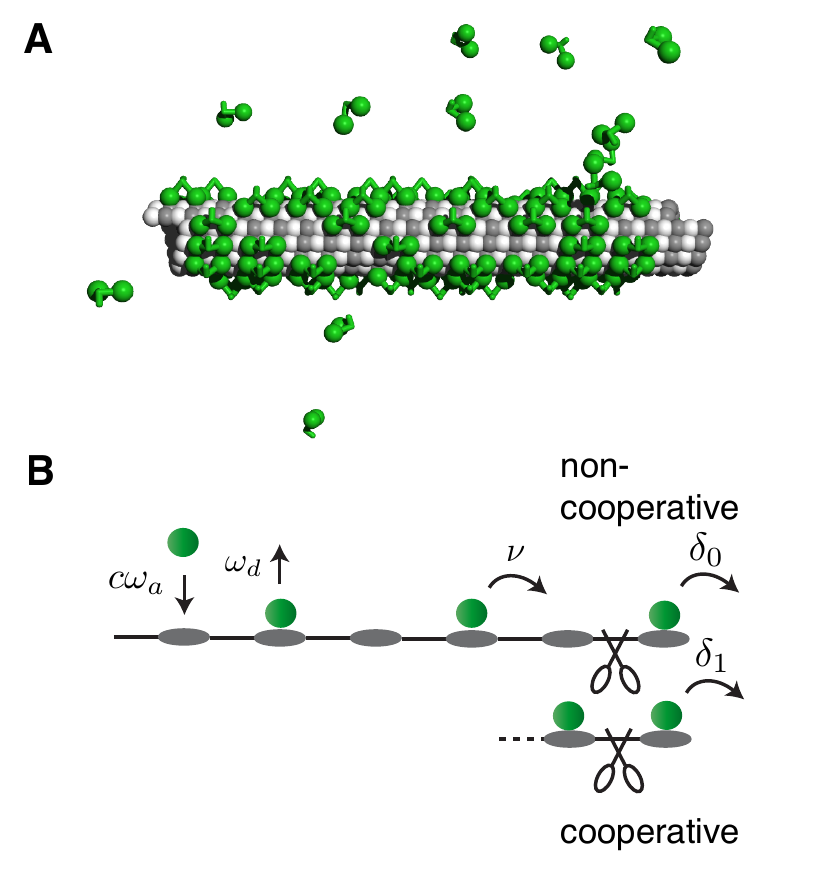}
\caption{\emph{Illustration of MT and motor dynamics.} Molecular motors present at concentration~$c$ randomly attach to unoccupied tubulin dimers along the MT lattice with rate $\omega_a$. While bound they processively move toward the plus-end at rate $\nu$, and unbind with rate $\omega_d$. Because motors do not switch lanes (protofilaments), the MT lattice ({\it A}) becomes effectively one-dimensional ({\it B}). Each lattice site $n_i$ (with $i=1, \ldots, L$ numbering the sites) may be  empty ($n_i = 0$) or occupied by a single motor ($n_i = 1$). At the plus-end the motors act as depolymerases (indicated by scissors) either alone with rate $\delta_0$ or cooperatively with rate $\delta_1$.
}
\label{fig:cartoon}
\end{figure}

We use an individual-based model, as illustrated in Fig.~\ref{fig:cartoon}, to describe the dynamics of plus-end-directed depolymerases. Motor proteins, present at a constant bulk concentration $c$, are assumed to randomly bind to and unbind from the MT lattice with rates $\omega_a$ and $\omega_d$, respectively. Bound motors are described as Poisson steppers (A more detailed biochemical model for motors on microtubules has to await further experimental analysis. One of the different possible schemes has recently been studied by Klumpp et al.~\citep{KLU08}.) that processively walk along individual protofilaments towards the plus-end at an average speed $v$~\citep{HOW96}. These motors hinder each other sterically because individual binding sites $i = 1, \ldots, L$ on each protofilament can either be empty ($n_i = 0$) or occupied by a single motor ($n_i = 1$). Because switching between protofilaments is rare \citep{HOW96}, transport along each of the protofilaments can be taken as independent, and the model becomes effectively one-dimensional~\citep{RAY93} (Fig.~\ref{fig:cartoon}{\it B}). 
Models of this type were recently discussed as minimal models for intracellular transport~\citep{PAR03, PAR04, LIP01, KLU03}. In its given formulation, where the cytosol is considered as a homogeneous and constant reservoir of motors, it is equivalent to a driven lattice gas model known as the totally asymmetric simple exclusion process with Langmuir kinetics (TASEP/LK)~\citep{PAR03}. A central finding of this model is that the interplay between on-off (Langmuir) kinetics and directed transport along protofilaments can result in ``traffic jams'' in which the density profile of motors along a protofilament shows a sharp increase from a low-density to a crowded high-density regime~\citep{LIP01, PAR03}. Such and other crowding effects~\citep{PIE06,TEL09} are important for a molecular understanding of MT dynamics. Previous theoretical studies on this topic largely disregarded crowding effects or considered parameter regimes in which they are unimportant~\citep{GOV08, BRU09, HOU09}. Depolymerization, including crowding effects, has also been investigated for diffusive depolymerases such as MCAK~\citep{KLE05}. 

At the plus-end of the system we consider depolymerization dynamics arising due to the interaction of molecular motors with the MT tip. Motivated by recent experiments~\citep{VAR09}, we assume  \emph{non-processive} depolymerization, i.e, a molecular motor dissociates from the lattice after triggering depolymerization. Because the molecular mechanisms are not yet fully resolved, we study two scenarios of depolymerization (see Fig.~\ref{fig:cartoon}{\it B}). In the noncooperative scenario, the dissociation rate depends only on whether the last site is occupied by a motor. If the last site is occupied, $n_L=1$, the MT depolymerizes at rate $\delta_0$. However, recent single molecule studies indicate that Kip3p may act \emph{cooperatively}~\citep{VAR09}, which we consider as our second scenario. After arriving at the plus-end, the motor is observed to pause and depolymerize a tubulin dimer only after a second Kip3p has arrived behind it. In this scenario, a tubulin dimer is depolymerized with rate $\delta_1$ if both the last and the second-to-last sites are occupied, $n_{L-1}=n_L=1$. Therefore, the total depolymerization rate can be written as:
\begin{equation}
\Delta = \delta_0 n_L + \delta_1 n_{L-1}  n_L \, . 
\end{equation}
For stabilized MTs, the spontaneous depolymerization rate is small~\citep{VAR09} and thus is not considered here. The relative magnitude of the noncooperative rate $\delta_0$ and the cooperative rate $\delta_1$ determines the degree of cooperativity of the depolymerization kinetics. In an average over many realizations of the stochastic process (ensemble average), the depolymerization speed $V_\text{depol}$ depends on the occupation of the last two binding sites by depolymerases (Fig.~\ref{fig:cartoon}{\it B}):
\begin{equation}
V_\text{depol} =  \left( \delta_0  \rho_+ + \delta_1 \kappa_+ \right) a \, ,
\label{eq:depol_speed}
\end{equation}
where $a$ is the lattice spacing. Here $\rho_+ := \avg{n_L}$ is the probability that the last site is occupied (i.e., the expected motor density at the plus-end), and $\kappa_+ :=  \avg{n_{L-1} n_L}$ denotes the probability that both the last and second-to-last sites are occupied. We analyzed this model via stochastic simulations and analytic calculations (for further details, see the Supporting Material).

\subsection*{\bf Validation of the model and its parameters}

The model parameters are, as far as they are available, fixed by experimental data. The motor speed, $v$, the motor run length, $\ell$, and motor association rate, $\omega_a$, were measured previously~\citep{VAR09}:
\begin{eqnarray}
v&=&3.2\,\mu\text{m}\,\text{min}^{-1}\, , \nonumber \\
\omega_a &=& 24\,\text{nM}^{-1}\text{min}^{-1}\mu\text{m}^{-1}\,, \nonumber \\
\ell&\approx&11\,\mu\text{m}\,. \nonumber
\end{eqnarray}
Using an MT lattice spacing of $a = 8.4\,\text{nm}$, we derive the corresponding parameters in our model as follows: The motor speed $v$ corresponds to $6.35$ lattice sites per second, i.e., a hopping rate of $\nu= v / a = 6.35\,\text{s}^{-1}$. The inverse hopping rate $\tau:=\nu^{-1}=0.16\,\text{s}$ and the size $a$ of a tubulin dimer serve as our basic timescale and length scale, respectively. Then, the measured association rate corresponds to a rate $\omega_a\approx 5.3\times 10^{-4}\,\text{nM}^{-1}\text{site}^{-1}\,\tau^{-1}$. The dissociation rate, $\omega_d = v /\ell $, is derived as the ratio of the mean motor speed, $v$, and the mean motor run-length, $\ell$. The latter equals $1310$ lattice sites. Thus, the dissociation rate is expressed as $\omega_d\approx7.6\times 10^{-4}\,\text{site}^{-1}\,\tau^{-1}$. In contrast to the transport behavior on the MT, the parameters concerning the depolymerization rates, $\delta_{0/1}$, cannot be directly extracted from experiments. However, there is evidence for a depolymerization rate as high as the motor speed, $v$~\citep{VAR06,VAR09}. As a starting point for the following discussion we tentatively take $\delta_{0}=\nu$.

\begin{figure}
\centering
\includegraphics*[width=6 in]{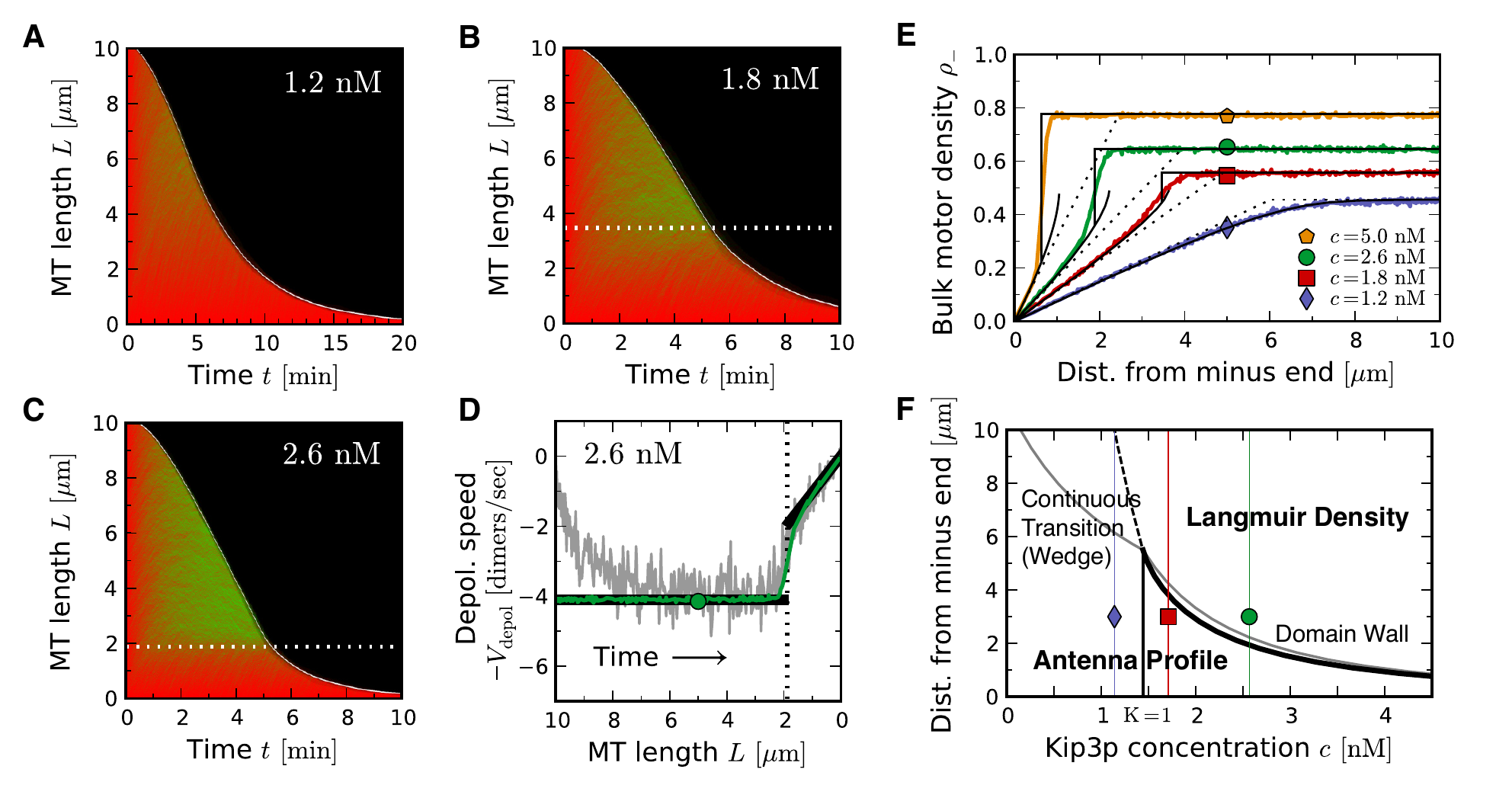}
\caption{\emph{Validation of the theoretical model.} ({\it A})-({\it C})~Time-space plots of stochastic simulations for a range of motor concentrations and depolymerization rate $\delta_0= 6.35\,\text{sites}\,\text{s}^{-1}$. The density of molecular motors is shown as the bright area ({\it green}), and the MT is shown as the dim area ({\it red}; for details, see Supporting Material). For low concentrations, $c < 1.4\,\mathrm{nM}$, depolymerization slows down gradually~\citep{VAR09}. At higher concentrations, $c > 1.4\,\mathrm{nM}$, there is a rather abrupt change in MT shortening. This change is correlated with a steep decrease in the motor density (DW), indicated as dotted lines. ({\it D})~The depolymerization speed, $V_\text{depol},$ as a function of the length of the MT $L(t)$, extracted from the simulation data shown in the kymograph ({\it gray}). The position of the DW ({\it dotted}), and the predicted depolymerization speed, $V_\text{depol}=v\rho(L)$ (see also Eq.~\ref{eq:experiment_depol}), using the linear approximation for the motor density profile ({\it black}) and the density profile extracted from stochastic simulations ({\it green}), coincide very well with the observed depolymerization speed; $v = 6.35\,\text{sites}\,\text{s}^{-1}$ is the walking speed of the motors. ({\it E})~Density profiles at the minus-end from stochastic simulations ({\it lines with symbols}), exact solutions ({\it solid}) and linearized theory ({\it dotted}) are shown. ({\it F}) As a function of the motor concentration, $c$, and the distance from the minus-end, there are distinct types of density profiles. At motor concentration lower than $c = 1.4\,\mathrm{nM}$ ({\it thin black}), the density of motors along the MT is low and the profile is smooth. The Langmuir density is reached continuously after a certain MT length ({\it dashed}, numerical). At high concentrations, $c > 1.4\,\mathrm{nM}$ there are two regions along the MT separated by an intervening DW ({\it black}, exact; see Supporting Material): an approximately linear antenna profile and a flat profile (Langmuir density). Linear approximations for the continuous and the discontinuous transitions (Eq.~\ref{eq:dw_position_minus}) are shown as well ({\it gray}). Thin lines refer to the density profiles shown in {\it E}.}
\label{fig:kymograph}
\label{fig:profile_diagram}
\end{figure}

Using the above set of parameters we now phenomenologically compare the results from numerical simulations of our model with observations from experiments. Specifically, we consider kymographs of the MT, which show how the MT length and the motor density on the MT evolve over time. For the simulation data shown in Fig.~\ref{fig:kymograph} we consider an MT consisting of 14 independent protofilaments and investigate the dynamics for the noncooperative scenario and a range of motor concentrations, $c = 1.2, \,1.8, \, 2.6 \, \text{nM}$, cf. Fig.~\ref{fig:kymograph}{\it A}-{\it C}. Surprisingly, as shown later, neither the cooperativity of the motors nor a decrease in the depolymerization rates led to different shapes of kymographs (see also Fig.~S1). We find an initial time period in which, starting from an empty MT lattice, the motors first fill up the lattice~\citep{VIL01, FRE02}. This is followed by a time window in which the motor density exhibits a quasi-stationary profile, i.e., the density at a certain distance from the minus-end does not change except for boundary effects induced by the plus-end. The corresponding density profiles are illustrated in Fig.~\ref{fig:kymograph}{\it E} and discussed in more detail in the following section. In this quasi-stationary regime, the depolymerization dynamics shows qualitatively different behavior depending on the concentration of free motor molecules: At low concentration, $c < 1.4\,\text{nM}$, and thus low density of motors on the MT, depolymerization slows down gradually in the course of time (Fig.~\ref{fig:kymograph}{\it A}). When the motor concentration increases to larger values, $c>1.4\,\text{nM}$, an intermediate regime emerges in which the depolymerization speed stays roughly constant (Fig.~\ref{fig:kymograph}{\it B} and {\it C}). Remarkably, we find that during this regime, the depolymerization speed is directly proportional to the motor density, $V_\text{depol} (L) = \rho_-(L) v$ (Fig.~\ref{fig:kymograph}{\it D}). At a third stage in the depolymerization process, there is a rather abrupt change in the depolymerization speed right where the density profile shows a steep drop (Fig.~\ref{fig:kymograph}{\it C}-{\it E}). After we have elaborated more on the theoretical model, we discuss why there is such a tight correlation between the depolymerization dynamics and the density profile.

All of these qualitative features of MT dynamics are identical to those found experimentally~\citep{VAR06,VAR09}, and suggest that the density profile and, in particular, traffic jams formed on the MT lattice are the main determinants of the depolymerization dynamics. Moreover, the time scales of the dynamics agree quantitatively well with experimental results for the same motor concentrations~\citep{VAR06,VAR09}. This validates our theoretical model because up to the depolymerization rate $\delta$, all of the model parameters were derived from experimental data~\citep{VAR09}. 

\subsection*{\bf Density profiles at the minus-end (bulk density)}

The above observations strongly point toward a tight correlation between the depolymerization speed and the motor density profile at the minus end, $\rho_-(x)$, which we henceforth call the \emph{bulk (motor) density}. The quasi-stationary bulk density profiles shown in Fig.~\ref{fig:kymograph}{\it E} were obtained by assuming very long lattices; effects caused by the plus-end are not visible in the vicinity of the minus-end. A more detailed discussion of these simulations can be found in the Supporting Material. Since this bulk density will play an important role in the following analysis, we summarize its features as obtained from analytical calculations detailed in the Supporting Material.  

\begin{figure}[t]
\centering
\includegraphics*[width=3.25in]{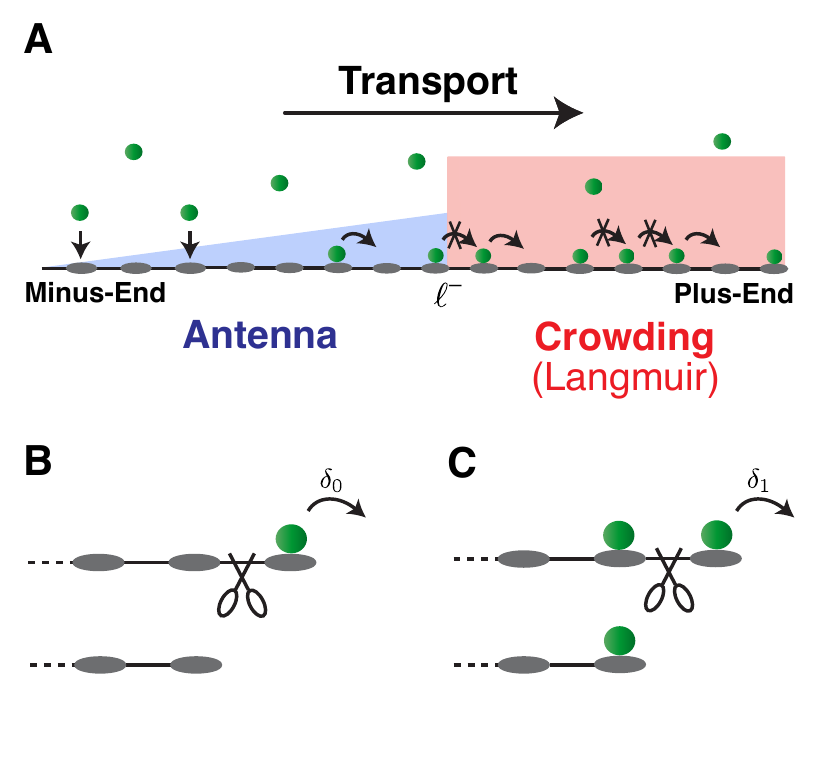}
\caption{\emph{Illustration of the antenna and crowding regimes, and of cooperativity}. ({\it A})~Starting from an empty MT, motors start to accumulate on the MT lattice by attachment and subsequent transport to the plus-end. The combined effect of Langmuir kinetics and steric exclusion between the motors leads to two sharply separated regimes. Starting from the minus-end, the motor density increases linearly (antenna profile). At a certain critical length $\ell^{-}$, a macroscopic traffic jam arises because particles hinder each other and crowding dominates the MT density. ({\it B})~Illustration of non-cooperative ({\it B}, nc) and fully cooperative ({\it C}, fc) depolymerization kinetics. With regard to the the depolymerization speed, both models are effectively equal (see main text).
}
\label{fig:illustration_crowding}
\label{fig:cartoon_depolymerization_kinetics}
\end{figure}

At the minus-end, the density profiles show an initial linear increase. This is an ``antenna effect''~\citep{VAR06} as illustrated in Fig.~\ref{fig:illustration_crowding}{\it A}. Motors that attach in proximity of the MT minus-end immediately move toward the plus-end, thereby generating an approximately linearly increasing accumulation of motors. The slope is given by $K/\ell$, where $K= {c \, \omega_a}/{\omega_d}$ denotes the binding constant. At sufficiently large distances from the minus-end, the density profile becomes flat and dominated by Langmuir kinetics with the ensuing Langmuir density:
\begin{eqnarray}
 \rho_\text{La} = \frac{K}{1+K} = \frac{c \, \omega_a}{c \, \omega_a + \omega_d} \, . 
 \label{eq:langmuir-vs-bulk}
\end{eqnarray}
The full density profile is obtained by concatenating the antenna profile and the flat Langmuir profile such that the motor current is continuous along the MT. We find two qualitatively distinct scenarios (Fig.~\ref{fig:kymograph}{\it E}). For low concentrations of molecular motors, $c$, the antenna profile matches the asymptotic Langmuir density continuously, resulting in a \emph{wedge-like} profile. In contrast, above a certain threshold value for the concentration, determined by the binding constant $K_c^-=1$, the two profiles can no longer be matched continuously and the density profile displays a sharp discontinuity, also termed a ``domain wall'' (DW)~\citep{PAR03}. In other words, if the Langmuir density rises above a critical value of $\rho_\text{La}^c = 0.5$, a crowding-induced traffic jam will result~\citep{FRE04} (Fig.~\ref{fig:illustration_crowding}{\it A}). The density profiles obtained from the analytic calculations and the stochastic simulations agree nicely, as illustrated in Fig.~\ref{fig:kymograph}{\it E}. In particular, the theoretical analysis gives an explicit expression for the width of the antenna-like profile:
\begin{equation}
\ell^- \approx \ell \, 
\begin{cases} 
\frac{1}{1+K} & \text{for} \; K<1 \, , \cr
 \frac{1}{K(1+K)}  & \text{for} \; K>1 \, .
\end{cases}
\label{eq:dw_position_minus}
\end{equation}
This result reduces to the average run length of molecular motors, $\ell = v/ \omega_d$, in the limit of very low binding constant, $K\ll1$, where crowding effects can be neglected~\citep{HOU09}. However, with increasing $K$ the regime with an antenna-like profile becomes significantly shorter than $\ell$ (Fig.~\ref{fig:profile_diagram}{\it F}).

\subsection*{\bf Depolymerization dynamics is independent of cooperativity}

We now address how the \emph{cooperativity} of the depolymerization kinetics affects the macroscopic depolymerization speed. There are two limiting cases: noncooperative depolymerization (nc) with $(\delta_0, \delta_1) = (\delta, 0)$, and fully cooperative depolymerization (fc) with $(\delta_0, \delta_1) = (0, \delta)$ (for an illustration, see Fig.~\ref{fig:cartoon_depolymerization_kinetics}, {\it B} and~{\it C}). Remarkably, we find from our stochastic simulations, shown in Fig.~\ref{fig:scaling}, that there is no difference in depolymerization speed for these two limiting cases. Even when the depolymerization dynamics contains cooperative as well as noncooperative terms, we do not find any significant differences in the depolymerization speed (Fig.~\ref{fig:scaling}{\it B}).

This observation from our stochastic simulations can be explained by the following molecular mechanism:  Consider a model with \emph{fully cooperative} depolymerization kinetics. Then, after the first motor has arrived at the plus-end, the terminal site of the MT will remain occupied from that time on. Depolymerization only occurs if another motor arrives at the second-to-last site. In other words, while the last site remains occupied, the second-to-last site triggers the depolymerization. Hence, as far as the depolymerization speed is concerned, the fully cooperative model is identical to a noncooperative model with the same molecular rate $\delta$. In the noncooperative model the terminal tubulin dimer is removed at rate $\delta$ once a molecular motor has arrived at the last site (see Fig.~\ref{fig:cartoon_depolymerization_kinetics}{\it B}). In the fully cooperative model, the terminal tubulin dimer is removed once a molecular motor has arrived at the second-to-last site next to a permanently occupied last site (Fig.~\ref{fig:cartoon_depolymerization_kinetics}{\it C}).

\subsection*{\bf Depolymerization dynamics is strongly affected by crowding}

\begin{figure}
\centering
\includegraphics*[width=3.25 in]{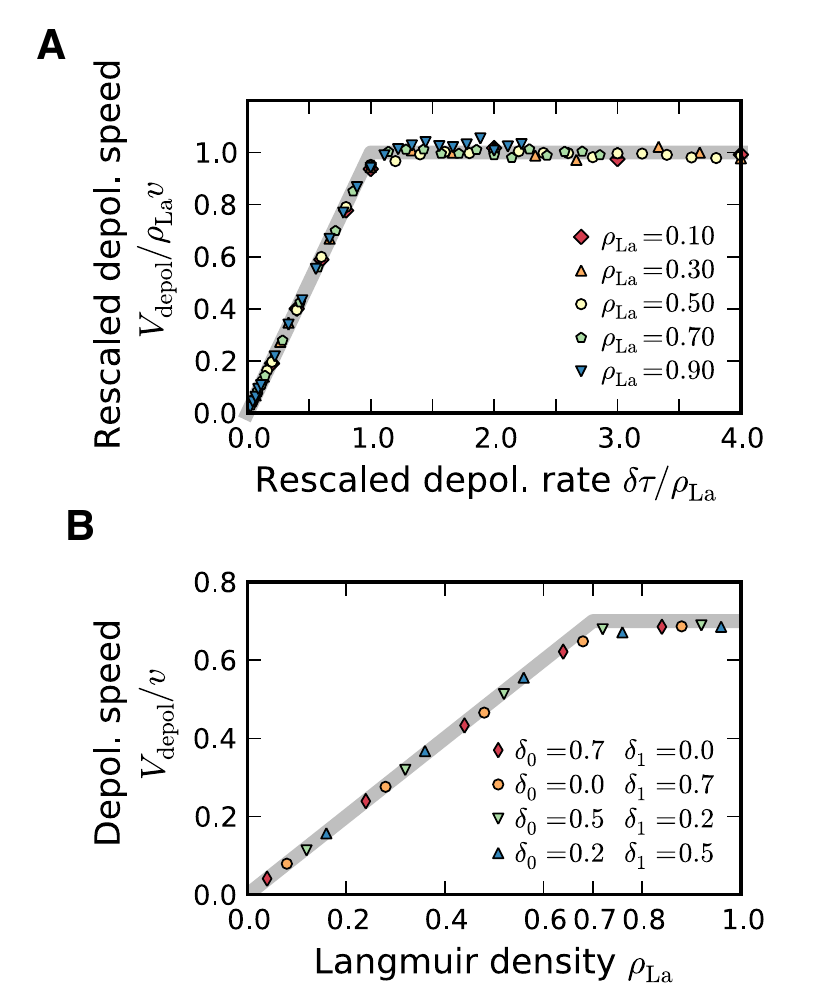}
\caption{\emph{Scaling plot for the depolymerization speed} $V_\text{depol}$.
({\it A}) Upon rescaling, both the macroscopic depolymerization speed, $V_\text{depol}$, and the microscopic depolymerization rate, $\delta$, with the Langmuir density, $\rho_\text{La}$, all data collapse onto one universal scaling function ${\cal V}$ ({\it solid gray}). A sharp transition at $\delta \tau = \rho_\text{La}^*$ distinguishes the rate-limited regime from the density-limited regime. 
({\it B}) Comparison of cooperative and noncooperative depolymerization, with the macroscopic depolymerization speed $V_\text{depol}$ as a function of Langmuir density $\rho_\text{La}$. For $\delta:=\delta_0 + \delta_1 =0.7\,\nu$ different degrees of cooperativity are displayed as indicated in the graph.}
\label{fig:scaling}
\end{figure}

To gain further insights in the correlation between the depolymerization speed and the density of motors on the MT, we performed stochastic simulations focusing on the MT plus-end by regarding the dynamics in a co-moving frame. Instead of simulating the full-length MT with an antenna profile and a subsequent flat Langmuir density, we considered a reduced model in which the density at the left end is set equal to the Langmuir density $\rho_\text{La}$. For long MTs, the Langmuir density is always reached, so that the reduced system is fully equivalent to the original model. Our simulations show two clearly distinct regimes of depolymerization dynamics (Fig.~\ref{fig:scaling}): For small microscopic depolymerization rates, $\delta \tau <\rho_\text{La}$, the depolymerization speed is \emph{rate-limited}: $V_\text{depol}= a \delta$. In contrast, for rates $\delta \tau >\rho_\text{La}$, the depolymerization speed is \emph{density-limited}, and the Langmuir density is the limiting factor: $V_\text{depol}=\rho_\text{La} v$. The boundary between the two regimes is remarkably sharp and given by 
\begin{equation}
\rho_\text{La}^* = \delta \tau \, .
\end{equation} 
This implies that the depolymerization speed can switch between being density-limited and rate-limited by changing the concentration $c$ or the values of the biochemical rates of depolymerases binding to and unbinding from the MT lattice. Overall, the depolymerization speed obeys a scaling law
\begin{eqnarray}
 V_\text{depol} = \rho_\text{La} v \, {\cal V} (\delta \tau/ \rho_\text{La}) 
 = \begin{cases}
    a \delta & \text{for} \; \delta \tau \leq \rho_\text{La} \cr
    \rho_\text{La} v & \text{for} \; \delta \tau > \rho_\text{La} 
    \end{cases}
    \, ,
\label{eq:speed_scaling_law}    
\end{eqnarray}
where ${\cal V} (x)$ is a universal scaling function with the simple form ${\cal V} (x) = x$ for $x<1$ and ${\cal V} (x) = 1$ for $x>1$. Experimentally, this implies that one should find data collapse upon using such a scaling plot (Fig.~\ref{fig:scaling}{\it A}). \\

\begin{figure}
\centering
\includegraphics*[width=3.25 in]{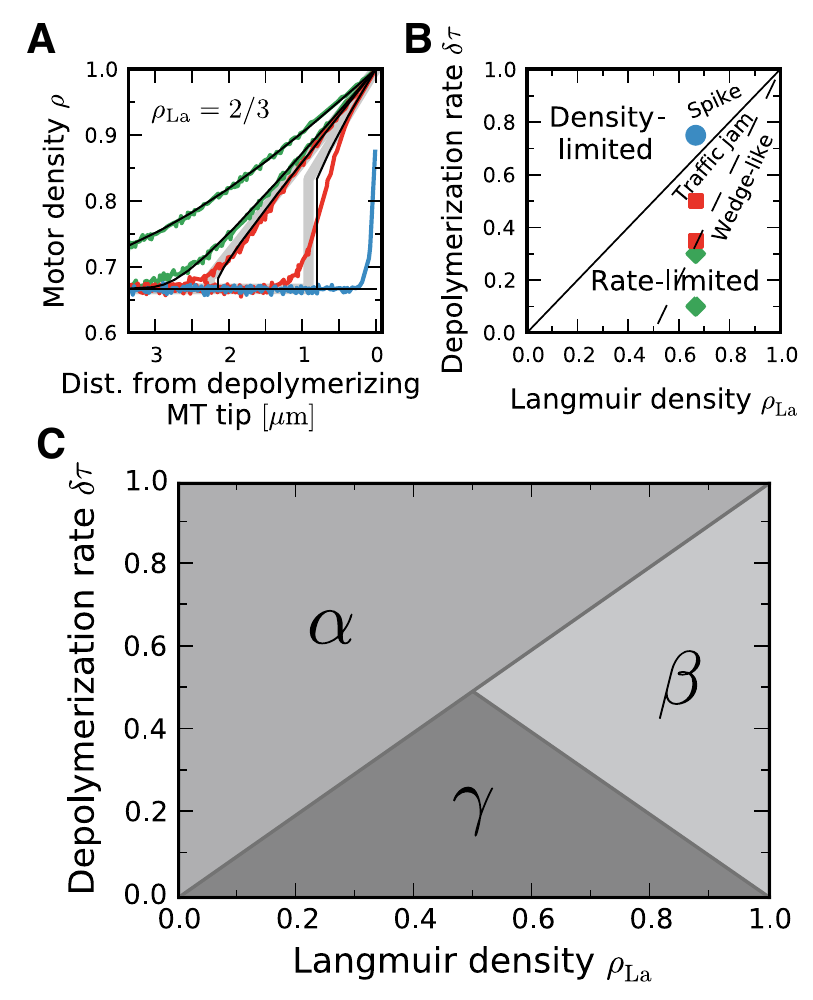}
\caption{\emph{Density profiles at the plus-end, corresponding phase diagram, and depolymerization scenarios.} ({\it A}) Density profiles at the MT plus-end in the comoving frame for $c=2.9\,\text{nM}$, and $\delta = 0.1, 0.3$ ({\it left}), $0.35, 0.5$ ({\it middle}) and $0.8\,\nu$ ({\it right}). The simulation results and analytical solutions ({\it black}; see Supporting Material) agree nicely.
({\it B}) Depending on the value of $\delta$ and the density of motors, $\rho_\text{La}$, there are three different classes of density profiles at the \emph{plus-end}: wedge-like ({\it diamonds}), traffic jams with a DW ({\it square}), and spikes ({\it circles}). The transition between profiles with an extended traffic jam and a localized spike ({\it solid line}) also marks a qualitative change in the depolymerization speed. Whereas the depolymerization speed is density-limited in the spike regime, it is rate-limited in the DW and wedge regime. Symbols correspond to parameters as displayed in panel {\it A}.
({\it C}) Depending on the value of $\delta$ and the density of motors, $\rho_\text{La}$, there are three different regimes of depolymerization dynamics. In regime $\alpha$ depolymerization is density-limited for arbitrary MT length. In contrast, depolymerization is rate-limited for long MTs and density limited for short MTs in the regimes $\beta$ and $\gamma$. For details, see the main text.}
\label{fig:tipdensities_fm}
\label{fig:phase_diagram}
\end{figure}

To gain a molecular understanding of these remarkable features of the depolymerization speed, one needs to have a closer look at the density profile of the molecular motors at the MT tip. If the depolymerization rate is small, $\delta<\nu$, motors leave the tip more slowly than they arrive. Therefore, the MT tip acts as a \emph{bottleneck} for molecular transport that disturbs the density profiles either locally or macroscopically. A weak bottleneck induces a \emph{local perturbation} (``spike'')~\citep{PIE06}. These spikes are sharp changes of the density profile with a typical extension that scales with the size of a heterodimer. However, if the strength of a bottleneck exceeds a threshold value, the spike extends to a \emph{macroscopic perturbation} (``traffic jam'')~\citep{PIE06}. 
Fig.~\ref{fig:tipdensities_fm}{\it A} illustrates how, for a given Langmuir density, $\rho_\text{La} = 2/3$, the effect on the density profile changes from a spike ({\it blue}) to an extended traffic jam  ({\it red} and {\it green}) when the depolymerization rate is $\delta$. 

Let us now analyze the conditions and consequences of such bottlenecks in more detail. Suppose we are in a parameter regime where the plus-end disturbs the density profile only locally, i.e., on the scale of a heterodimer. Then, we may take the bulk density to be equal to the Langmuir density, $\rho_\text{La}$, up to the last site (the plus-end) where it jumps to some higher or lower value $\rho_+$. The particle loss current at the plus-end due to MT depolymerization is then given by
\begin{equation}
J_\text{depol} = (1-\rho_\text{La}) \rho_+ \delta \, .
\label{eq:loss}
\end{equation}
The factor $1-\rho_\text{La}$ arises because the particle number decreases only if a particle depolymerizes the MT and the second-to-last site, $L-1$, is unoccupied. Otherwise, depolymerization dynamics and the associated frame shift of the MT lattice do not change the occupation of the last site. This particle loss has to be balanced by the incoming particle flux, 
\begin{equation}
J_\text{La} = \rho_\text{La}(1-\rho_\text{La}) \nu\,.
\label{eq:gain}
\end{equation}
Equating these particle fluxes (Eqs.~\ref{eq:loss} and \ref{eq:gain}) implies the following condition for the motor density at the plus-end: 
\begin{equation}
\rho_+=\begin{cases}
 \rho_\text{La} / {\delta \tau} & \text{for} \; \rho_\text{La} \leq \delta \tau \cr
1 & \text{for} \; \rho_\text{La}>\delta \tau
    \end{cases},
\label{eq:rho+}
\end{equation}
where the fact that the motor density is bounded $\rho_+\leq 1$ is already accounted for. The particle density on the last site, in turn, determines the depolymerization speed. For $\rho_\text{La} < \delta \tau$, one obtains according to Eq.~\ref{eq:depol_speed} and Eq.~\ref{eq:rho+}:
\begin{equation}
V_\text{depol} = \rho_+ \, \delta a = \rho_\text{La} v \, .
\label{eq:experiment_depol}
\end{equation}
Remarkably, here the effect of the depolymerization kinetics ($\delta$) cancels out such that the macroscopic depolymerization speed is independent of the molecular details of depolymerization kinetics and solely determined by the Langmuir density, i.e., the motor density in the bulk, $\rho_- (x)$, and \emph{not} at the tip of the MT. This result crucially depends on the presence of a microscopic spike. It explains the hitherto puzzling experimental result that the depolymerization speed is directly proportional to the bulk motor current along the MT~\citep{VAR09} (Fig.~S2). 

Because the density is bounded, $\rho_+ \leq 1$, density profiles with a spike are only possible if the densities are not too large, $\rho_\text{La} < \delta \tau$. This is the case for the blue curve in Fig.~\ref{fig:tipdensities_fm}{\it A}. For densities exceeding the critical density, $\rho^*_\text{La} = \delta \tau$, the bottleneck-induced perturbation in the density profile can no longer remain a local spike, but has to become macroscopic in extent~\citep{PIE06} (see {\it green and red curves} in Fig.~\ref{fig:tipdensities_fm}{\it A} and Supporting Material).
One finds that over an extended region, the binding sites at the plus-end then remain permanently occupied such that $\rho_+=1$. This immediately implies that the depolymerization speed becomes density-independent and proportional to the microscopic depolymerization rate:
\begin{equation}
V_\text{depol} = a \delta\, .
\end{equation}
There is a tight correlation between the shape of the density profiles and the macroscopic depolymerization speed. The analytic results explain the molecular mechanism behind the numerically observed scaling law (Eq.~\ref{eq:speed_scaling_law}), with a sharp transition from density-regulated to a rate-limited depolymerization dynamics at a critical value of $\rho_\text{La}^*=\delta \tau$ (cf. the classification of density profiles and depolymerization regimes shown in  Fig.~\ref{fig:phase_diagram}{\it B}).

Actually, the above calculations can be generalized to the regime in which the motor density exhibits an antenna-like linear profile, i.e., for MT length shorter than $\ell^-$. As detailed in the Supporting Material, we find that the depolymerization speed is rate-limited, $V_\text{depol}= a \delta$, if MTs are shorter than $\ell^-$ but still longer than a second threshold length: 
\begin{equation}
\ell_d := \delta a / c \omega_a =  \ell \, \delta \tau / K \, .
\end{equation} 
In contrast,  for $\ell_d > \ell^-$, the depolymerization speed in the antenna regime is \emph{always} length-dependent and strictly follows the shape of the antenna profile, $\rho_- (x)$:
\begin{equation}
V_\text{depol} = \rho_- (L) v \, .
\end{equation}
Using Eq.~\ref{eq:dw_position_minus}, the condition $\ell_d > \ell^-$ on the threshold lengths is equivalent to $\delta \tau  > \rho_\text{La}$ for $K<1$ and to $\delta \tau  > 1- \rho_\text{La}$ for $K>1$.

Combining all of the above results, we find three mechanisms governing depolymerization dynamics, as illustrated in Fig.~\ref{fig:phase_diagram}{\it C}: 
\begin{itemize} 
\item [($\alpha$)] For $\delta \tau  > \rho_\text{La}$, the depolymerization speed is always density-regulated and given by $V_\text{depol} (L) = \rho_- (L) v$, where $L$ is the time-dependent length of the MT. In this parameter regime, the depolymerization speed is a direct map of the bulk motor density profile on the MT, $\rho_- (x)$, a feature that can be exploited experimentally to measure the profile. 
\item [($\beta$)] For $\rho_\text{La} > \delta \tau > 1- \rho_\text{La}$, the depolymerization speed is rate-limited for MTs longer than $\ell^-$, and becomes density-limited as soon as the MT length falls below $\ell^-$ where the density profile is antenna-like. This implies that there is a \emph{discontinuous} jump in the depolymerization speed right at $L=\ell^-$. 
\item [($\gamma$)] Finally, for all other values for $\delta \tau$, the depolymerization speed of the MT remains rate-limited for lengths larger than a threshold length $\ell_d$. At $\ell_d$, which is smaller than $\ell^-$ in this parameter regime, there is again a discontinuous jump to a density-limited depolymerization dynamics. 
\end{itemize}  

If the depolymerization rate is larger or equal to the hopping rate of molecular motors, $\delta \tau \geq 1$, then $\delta \tau  > \rho_\text{La}$ is always obeyed simply because $\rho_\text{La} \leq 1$. In this regime, all of the molecular details of the depolymerization kinetics are irrelevant. Neither cooperativity nor the actual value of the depolymerization rate matters in terms of the depolymerization speed; instead, only the bulk density regulates the speed. Note that this was the case for the data shown in Fig.~\ref{fig:profile_diagram}, where we tentatively made the parameter choice $\delta\tau = 1$. If the motors are faster than the depolymerization process, $\delta \tau < 1$, we have to distinguish between the parameter regimes ($\alpha$,$\beta$, and $\gamma$, Fig.~\ref{fig:phase_diagram}{\it C}). Here the value of the depolymerization rate matters if the bulk density exceeds a certain threshold concentration, $\rho_\text{La} > \delta \tau$, and the MTs are long enough. Finally, the depolymerization speed always becomes density-dependent and hence length-dependent if the MT length is short enough; the corresponding threshold length is $\ell_\text{reg} = \text{min} [\ell^-, \ell_d]$.  

\subsection*{\bf End-residence time strongly depends on cooperativity}

\begin{figure}
\centering
\includegraphics*[width=3.25 in]{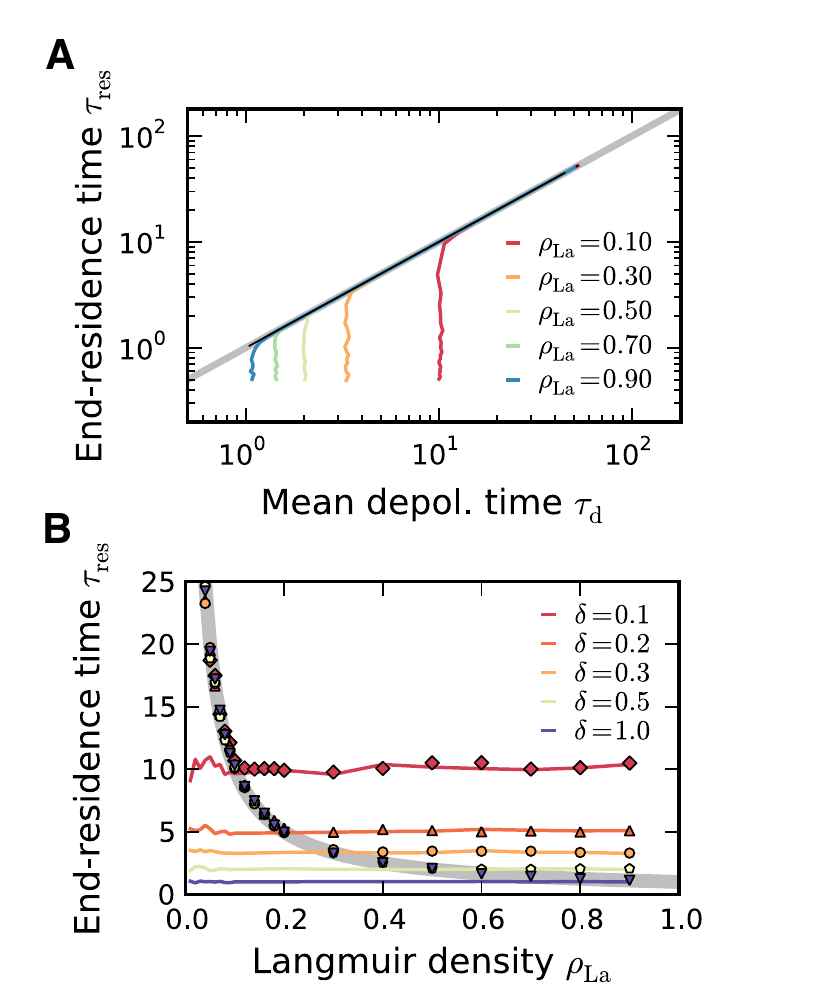}
\caption{\emph{Motor end-residence times} $\tau_\text{res}$ \emph{for cooperative and noncooperative depolymerization}.
({\it A}) Mean end-residence time $\tau_{\text{res}}$ plotted against the mean depolymerization time $\tau_d$. Data were recorded for a range of depolymerization rates $\delta=0.02\ldots2\,\nu$. Noncooperative ({\it shaded}) and cooperative ({\it black}) dynamics are shown for different densities. 
({\it B}) Mean end-residence time $\tau_{\text{res}}$ as a function of the Langmuir density $\rho_\text{La}$ for various depolymerization rates (in units of $\nu$). For noncooperative depolymerization, $\tau_{\text{res}}$ is given by $1/\delta$ ({\it shaded lines}). For the fully cooperative scenario ({\it symbols}), $\tau_{\text{res}}$ depends on whether the system is in the density-limited ($\delta \tau>\rho_\text{La}$) or in the rate-limited ($\delta \tau <\rho_\text{La}$) regime. While, for $\delta \tau >\rho_\text{La}$, the end-residence time is given by $\tau_{\text{res}}= \tau / \rho_\text{La}$ ({\it solid gray line}), for $\delta \tau <\rho_\text{La}$, it is density-independent and determined by the microscopic depolymerization rate $\tau_{\text{res}}=1/\delta$ (see also Eq.~\ref{eq:taures}).
}
\label{fig:end_residence_time}
\end{figure}

In contrast to the depolymerization speed, the mean end-residence time $\tau_\text{res}$ is strongly affected by the degree of cooperativity. Fig.~\ref{fig:end_residence_time} displays $\tau_\text{res}$ as obtained from our stochastic simulations for noncooperative and fully cooperative depolymerization kinetics. Our simulations show that the end-residence time for the fully cooperative model is identical to the average lifetime of a terminal tubulin dimer $\tau_\text{res}^{\text{fc}}=\tau_d := a / V_\text{depol}$ (Fig.~\ref{fig:end_residence_time}{\it A}). Even for the noncooperative model, $\tau_\text{res}^\text{nc}$ equals $\tau_d $ for large residence times and deviates from it only at small values. The relatively sharp transition to a constant lifetime of the terminal tubulin dimer occurs right at $\tau_\text{res}^\text{nc} = \tau / \rho_\text{La}$, i.e., the end-residence time equals the waiting time for a molecular motor to arrive at the MT tip. For $\tau_\text{res}^\text{nc}  < \tau / \rho_\text{La}$, the lifetime of the terminal tubulin dimer is identical to the arrival time (Fig.~\ref{fig:end_residence_time},{\it A} and~{\it B}).
Once the arrival time becomes shorter than the inverse depolymerization rate, the end-residence time levels off at $\tau_\text{res} ^\text{nc} = 1 / \delta$. These results show that the dependence of the end-residence time on density can be used to quantify the degree of cooperativity. This would require experiments with motor densities on the MT larger than those studied up to now~\citep{VAR06, VAR09}. 

The observation that the depolymerization speed is independent of the degree of cooperativity seems to be at odds with the experimental finding that the end-residence time, $\tau_\text{res}$, of Kip3p depends on the total Kip3p concentration and is inversely proportional to the macroscopic depolymerization speed \citep{VAR09}. Actually, however, there is no contradiction and the findings are readily explained within our theoretical model: For a noncooperative model, $\tau_\text{res}^\text{nc} $ is simply given by the depolymerization rate, because after they arrive, the particles stay at the tip until they depolymerize the MT:
\begin{equation}
\tau_\text{res}^\text{nc} = \frac{1}{\delta} \, .
\end{equation}
For a fully cooperative model, $\tau_\text{res}^\text{fc} $ depends not only on $\delta$, but also on the rate at which the second-to-last site becomes populated. Say the probability for the second-to-last site to be occupied is $\rho_+$. Then, $\tau_\text{res}^\text{fc} $ is given by a sum of two contributions arising from the cases in which the second-to-last site is empty or occupied, respectively:
\begin{equation}
\tau_\text{res}^\text{fc} = (1-\rho_+) \left( \frac{\tau}{\rho_\text{La}} + \frac{1}{\delta} \right)
+  \rho_+ \,  \frac{1}{\delta}  \, .
\label{eq:tauresfc}
\end{equation}
If-the-second to last site is empty (which is the case with probability $1-\rho_+$) $\tau_\text{res}$ is the sum of arrival time $\tau/\rho_\text{La}$ and depolymerization time $1/\delta$. Otherwise, the end-residence time $\tau_\text{res}$ simply equals $1/\delta$.  

As shown in the previous section, two distinct scenarios arise: For small bulk densities such that $\rho_\text{La} < \delta \tau$, the density profile at the plus-end exhibits a microscopic spike with $\rho_+ = \rho_\text{La}/\delta \tau$. For large densities, $\rho_\text{La} > \delta \tau$, a macroscopic traffic jam emerges such that $\rho_+ = 1$.
This result obtained for the motor density at the MT tip (Eq.~\ref{eq:rho+}) may now be used to calculate $\tau_\text{res}^\text{fc}$ using Eq.~\ref{eq:tauresfc}: 
\begin{equation}
\tau_\text{res}^\text{fc} = 
\begin{cases}
\frac{1}{\delta} & \text{for} \; \rho_\text{La} > \delta \tau \, , \cr
\frac{\tau}{\rho_\text{La}} & \text{else} \, .
\end{cases}
\label{eq:taures}
\end{equation}
This agrees well with the results from stochastic simulations displayed in Fig.~\ref{fig:end_residence_time}. A comparison with Eq.~\ref{eq:speed_scaling_law} shows that the end-residence time equals the typical depolymerization time, i.e., the expected lifetime of a terminal tubulin dimer, $\tau_\text{res}^\text{fc} =\tau_d$. This is in agreement with  experimental findings regarding the unbinding-rate of motors at the plus-end~\citep{VAR09} and strongly supports the conclusion that depolymerization of MTs by Kip3p is fully cooperative. Varga et al.~\citep{VAR09} measured the end-residence time of motors on double stabilized MTs, i.e., where depolymerization is switched off. They observed that the end-residence time is inversely correlated with the concentration of Kip3p, and fit their data with an exponential using a cut-off. This is in accordance with our results shown in Fig.~\ref{fig:end_residence_time}{\it B}.
However, since depolymerization has been switched off in the experiment, the rate $\delta$, corresponding to the cutoff, now has to be interpreted as an unbinding-rate of motors at the plus-end.
It would be highly interesting to design experiments where the depolymerization kinetics remains switched on, because this would allow one to measure the magnitude of the microscopic depolymerization rate $\delta$.

\section*{\bf DISCUSSION}

In this work, we have analyzed the effect of crowding and cooperativity on the depolymerization dynamics of microtubules. To that end, we constructed an individual-based model for the coupled dynamics of plus-end directed motor traffic and microtubule depolymerization kinetics. The model is based on well-established molecular properties of motors from the kinesin-8 family, i.e., the motors move on single protofilaments with high processivity at an average speed $v$, and exchange of motors between the bulk and the microtubule follows Langmuir kinetics. All parameters of the model, including the average walking speed, run length, and attachment rate, were directly extracted from available \emph{in vitro} data~\citep{VAR09}. We have validated our model by reproducing the onset of length-dependent depolymerization as studied recently~\citep{VAR06,VAR09}. Without using any additional fitting parameter, we found the same regimes of density profiles and ensuing depolymerization dynamics as in the experiments, i.e., a linear antenna-profile with a length-dependent depolymerization speed and a flat profile with a constant depolymerization speed. Moreover, we identified a threshold density of motors above which a crowding-induced traffic jam emerges at the minus-end. The predicted shape and extent of these traffic jams should be amenable to experiments that raise the depolymerase concentration $c$ or changing its rates of binding to and unbinding from the MT.

The interplay between motor traffic and depolymerization kinetics at the microtubule plus-end leads to strong correlations between the depolymerization dynamics and density profiles of depolymerases. The plus-end acts as a bottleneck and crowding effects cause traffic jams. We find two qualitatively distinct regimes: Motor densities below a critical threshold value, $\rho_{La}^* = \delta \tau$, always show a local spike-like perturbation at the plus-end, the extent of which is the size of a heterodimer. Above this threshold density, macroscopic traffic jams may emerge. These distinct density profiles at the plus-end affect the depolymerization speed and the end-residence time in qualitatively different ways. A quantitative analysis of the model using stochastic simulations as well as analytical calculations led to the following main results:
The \emph{end-residence time} of a depolymerase strongly depends on the degree of cooperativity. Whereas for noncooperative depolymerization kinetics the end-residence time is given by the microscopic depolymerization rate $\delta$, it is density-dependent in the fully cooperative case: Increasing the Langmuir density above the threshold value $\rho_\text{La}^* = \delta \tau$, the end-residence time changes from being inversely proportional to the density $\rho_\text{La}$ to a constant value $\delta^{-1}$. These results suggest an interesting way to determine the cooperativity of depolymerization kinetics and measure the value of the depolymerization rate $\delta$. Although when the concentration $c$ is increased, the end-residence time should be independent of concentration for noncooperative kinetics, it should strongly depend on concentration in the cooperative case. Experimental evidence points toward the latter \citep{VAR09}. 

In contrast, the \emph{depolymerization speed} does not depend on the degree of cooperativity of the depolymerization kinetics. Noncooperative and fully cooperative versions of the model give identical results. As a function of depolymerase concentration and the MT length, the depolymerization dynamics exhibits two qualitatively distinct regimes: The depolymerization speed is either \emph{density-limited} and determined by the \emph{bulk density of molecular motors}, $\rho_- (x)$, or \emph{rate-limited} and dictated by the value of the \emph{microscopic depolymerization rate}, $\delta$. Both regimes emerge due to \emph{crowding of molecular motors} at the plus-end which acts as a bottleneck for molecular traffic. 

Density-limited regimes are correlated with microscopic traffic jams (``spikes'') at the plus-end: The density profile self-organizes into a shape that cancels out all the effects of the depolymerization kinetics such that the depolymerization speed is solely determined by the bulk motor density, $\rho_- (x)$,  and the average motor speed, $v$. Note that only in this regime length-dependent regulation is possible since the density changes over the MT length. As emphasized above, if the depolymerization rate $\delta$ is larger than the hopping rate of the molecular motors, $\delta  > \nu$, this remains the only regime of depolymerization dynamics. Then, the depolymerization speed is limited by the velocity of the plus-end directed motors, which is in accordance with recent experimental findings for Kip3p~\citep{VAR09}. 
In a parameter regime where motors depolymerize more slowly than they walk, $\delta < \nu$, there is a second \emph{rate-limited} regime above the threshold density $\rho_\text{La}^*$ and for microtubules longer than some threshold length $\ell_\text{reg}$ where $V_\text{depol} = a \delta$. 
In this regime the plus-end acts as a strong bottleneck for molecular traffic. This causes a macroscopic traffic jam such that the motor density steeply rises to full occupation of all lattice sites at the plus-end of the microtubule. The cellular system sacrifices its capability to regulate the speed of depolymerization and only regains it once the MT length falls below $\ell_\text{reg}$, where the depolymerization speed again becomes density-regulated. From an evolutionary perspective one might speculate that the system has evolved towards $\delta = \nu$, because this would allow regulation of the depolymerization dynamics over the broadest possible range.

Beyond these observations, other predictions of our stochastic model can be put to test in experiments. By varying the motor concentration, two interesting observations could be made: First the phase diagram for the density profiles at the minus-end could be scrutinized experimentally. Second, the predictions on the density-profiles at the plus-end and their predicted strong correlations to the macroscopic depolymerization dynamics might be accessible to single-molecule studies. Manipulation of the molecular properties of the motor (e.g., the run length, attachment rate~\citep{COO10}, average speed and depolymerization rate) would change the intrinsic biochemical rates of the system and potentially lead to new parameter regimes. In addition, our results regarding length- and concentration dependence of the depolymerization process might be relevant \emph{in vivo}, e.g., for mitotic chromosome alignment~\citep{STU08}. In our theoretical studies we explored the full parameter range, and therefore clear predictions are available for comparison. 

We believe that in a more general context, our theoretical work provides new conceptual insights into the role of collective and cooperative effects in microtubule assembly and disassembly dynamics. Future research could focus on the antagonism between polymerases and depolymerases~\citep{KIN01,HOW07,BRO08}, spontaneous MT dynamics mediated by GTP-hydrolysis, the abundance of molecular motors in a cell, or more-detailed modeling of molecular motors~\citep{KLU08}. This may finally lead to a molecular understanding of the regulatory mechanisms of cellular processes in which MT dynamics plays a central role.

\section*{\bf ACKNOWLEDGMENTS}
The authors thank C\'{e}cile Leduc for discussions, the authors of~\citep{VAR09} for kindly providing their data, Ulrich Gerland, G\"unther Woehlke and Jonas Cremer for critical reading of the original manuscript, Anton Winkler for helpful suggestions on the revised manuscript and Andrej Vilfan for drawing~Fig.~1{\it A}. This project was supported by the Deutsche Forschungsgemeinschaft in the framework of the SFB 863 and the German Excellence Initiative via the program ``Nanosystems Initiative Munich'' (NIM).

\newpage

\appendix

%\newcounter{Sequation}
\setcounter{equation}{0}
\newcounter{Sfigure}
\renewcommand{\theequation}{S\arabic{equation}}
\renewcommand{\thefigure}{S\arabic{Sfigure}}
\setcounter{Sfigure}{1}
\renewcommand{\figurename}{FIGURE}

%{ \huge \bf Supporting Material\\}
\begin{center}
\Large{Supporting Material to ``Crowding of molecular motors determines microtubule depolymerization''}
\vspace{.5 cm}\\
\normalsize Louis~Reese, Anna~Melbinger and Erwin~Frey \\(Corresponding author. Email:~frey@lmu.de) \vspace{.5cm}
\\ \footnotesize Arnold Sommerfeld Center for Theoretical Physics and Center for NanoScience, \\ Department of Physics,  Ludwig-Maximilians-Universit\"{a}t M\"{u}nchen, \\Theresienstra\ss e 37, 80333 Munich, Germany
\end{center}

In this Supporting Material, details concerning the mathematical formulation and the stochastic simulations are given. In particular,  the density profiles and the domain wall positions at the minus- and the plus-end are derived analytically. Further, some additional results are provided: (i) We show that the shapes of MT depolymerization curves (kymographs) are to a large extent independent of the choice of the depolymerization rate $\delta$; see  Fig.~\ref{fig:kymo}. (ii) Analytical and numerical results from our theory are compared to experimental data on the relation between depolymerization speed and motor current ~\citep{VAR09S}; see Fig.~\ref{fig:depolimerization_speed_current}. 

\section*{Mathematical formulation}
In this article, we employ a lattice gas model. Its state is described by a set of occupation numbers $n_i \in \{ 0,1 \}$ where $i=1, \ldots, L$ denotes the lattice sites.
In contrast to the notation in the main text, we here choose units of length and time such that the hopping rate from site to site and the lattice constant are both set to one. For an analytical description of the steady state density profiles of the molecular motors along the MT we consider the ensemble-averaged densities and currents:
\begin{eqnarray}
 \rho_i := \avg{n_i} \, ,\\
 J_i:= \avg{n_i (1-n_{i+1})} \label{fig:current_density_rel} \, .
\end{eqnarray}
Note that  the current $J_i$ accounts for particle exclusion: a particle at site $i$ moves to site $i+1$ at rate $\nu=1$ only if site $i+1$ is unoccupied. The steady state results from a local balance between the transport current~\citep{PAR03S},
\begin{eqnarray}
 J^T_i:= \avg{n_{i-1} (1-n_i)} -  \avg{n_{i} (1-n_{i+1})} \, ,
\end{eqnarray}
the particle exchange with the bulk,
\begin{eqnarray}
 J^{La}_i:= c \, \omega_a \avg{ 1 - n_{i} } - \omega_d \avg{ n_{i} }  \, ,
\end{eqnarray}
and the depolymerization current, which sets the boundary condition at the plus-end. We now perform a mean-field approximation, where all spatial correlations are neglected, and a continuum limit keeping only the leading order terms \citep{PAR04S}. Then, the transport current simplifies to,
\begin{eqnarray}
 J^T (x) = \bigl( 2 \rho(x)-1 \bigr) \partial_x \rho(x) \, , 
\end{eqnarray}
i.e. the transport current is proportional to the density gradient like a diffusion current in Fick's law but modified with a density-dependent prefactor which reflects site-exclusion between motors. The Langmuir current is given by 
\begin{eqnarray}
 J^{La} (x) = c \, \omega_a \bigl( 1 - \rho(x) \bigr) - \omega_d \rho (x) \, .
\end{eqnarray}

\section*{Density profiles at the minus-end}

Within the above introduced framework the motor density profiles on the MT can be calculated analytically. In particular, the domain wall position, can be derived exactly as well as upon employing a linear approximation for the density profile close to the minus-end. For simplicity, we first consider the latter, especially because its results approximate the exact solution rather well over a broad range of parameters.

\subsection*{Linear approximation}

In the immediate vicinity of the minus-end ($x=0$) the density is small such that the full equation for the current balance, $J^T+J^{La}=0$,
\begin{eqnarray}
(2 \rho(x) - 1) \partial_x \rho(x) + c \, \omega_a (1 - \rho(x)) - \omega_d \rho(x) = 0 \, ,
\label{eq:tasep_lk}
\end{eqnarray}
reduces to $\partial_x \rho =  c \, \omega_a$,  which is solved by a linear (antenna) profile: 
\begin{eqnarray}
\rho_- (x) \approx c \, \omega_a x \, .
\label{eq:tasep_linear}
\end{eqnarray}
At sufficiently large distances from the minus-end the density profile becomes flat.  Therefore, $J^T$ vanishes and the system is dominated by the Langmuir kinetics, $J^{La}=0$. Then, an asymptotic solution of Eq.~\ref{eq:tasep_lk} is given by the Langmuir density
\begin{eqnarray}
 \rho_\text{La} = \frac{K}{1+K} = \frac{c \, \omega_a}{c \, \omega_a + \omega_d} \, . 
 \label{eq:langmuir-vs-bulk_supp}
\end{eqnarray}
The full density profile is obtained by concatenating the antenna profile and the flat Langmuir profile such that the (local) current is continuous along the MT. There are two qualitatively distinct scenarios. For low bulk concentrations of molecular motors, $c$, the antenna profile matches the asymptotic Langmuir density continuously resulting in a \emph{wedge-like} profile; compare Fig.~2{\it E} in the main text. Approximately, the matching point, $\rho_- (d_w^-) = \rho_\text{La} $, is 
\begin{eqnarray}
 \ell_w^- \approx \frac{K}{c \, \omega_a (K+1)} \, . 
 \label{eq:dens_match_linear}
\end{eqnarray}
In contrast, above a certain threshold value for the bulk concentration, determined by $K_c^-=1$, the two profiles can no longer be matched continuously and the density profile displays a localized discontinuity~\citep{PAR03S}, also termed a \emph{``domain wall"} (DW). Its position is determined by a local current continuity condition~\citep{PAR03S,PAR04S}, $\rho_- (d^-) = 1 - \rho_\text{La}$, and can again be estimated using the linear antenna profile:
\begin{eqnarray}
\ell^- \approx \frac{1}{c \, \omega_a (K+1)} \, .
\label{eq:dw_position_linear}
\end{eqnarray}
Taken together, Eq.~(5) from the main text is obtained,
\begin{equation}
\ell^- = 
\begin{cases} 
\frac{1}{\omega_d (K+1)} & \text{for} \; K<1 \, , \cr
\frac{1}{c \, \omega_a (K+1)} & \text{for} \; K>1 \, .
\end{cases}
\label{eq:length_threshold}
\end{equation}

\subsection*{Exact solution and domain wall position}
To obtain the full solution Eq.~\ref{eq:tasep_lk} has to be solved as already demonstrated in Ref.~\citep{PAR04S}.
Introducing a rescaled density at the minus-end $\sigma_-(x)=\frac{K+1}{K-1}\left(2\rho(x)-1\right)-1$ in Eq.~\ref{eq:tasep_lk} a transformed differential equation can be obtained
\[
\partial_x\sigma_-(x) + \partial_x \ln{\left|\sigma(x)_-\right |} = \omega_d\frac{(K+1)^2}{K-1},
\]
which is mathematically equivalent to Eq.~\ref{eq:tasep_lk} and can be solved analytically
\begin{equation}
\sigma_-(x)=W_{-1}\left(-Y_-(x)\right).
\label{eq:sol_minus}
\end{equation}
Here $W_{-1}$ is the second real branch of the Lambert $W$-function~\citep{COR96S} and $Y_-(x)$ reads~\citep{PAR04S}
\begin{equation}
Y_-(x)=\left| \frac{-2K}{K-1}\right|\exp\left\{ \omega_d\frac{(K+1)^2}{K-1} x-\frac{2K}{K-1}\right\}.
\label{eq:sol_minus_y}
\end{equation}
Herein the boundary condition $\rho_-(0)=0$ corresponding to $\sigma_-(0)=-2K/(K-1)$ has already been accounted for. The local current condition for the domain wall, $\rho_-(d^-)=1-\rho_{La}$ which corresponds to  $\sigma_-(d^-)=-2$ for the rescaled density, now enables us to calculate the DW position. Combining this condition with Eqs.~\ref{eq:sol_minus} and \ref{eq:sol_minus_y} leads to
\begin{equation}
d^-(\omega_d,K)=\frac{2+(K-1)\ln(1-1/K)}{\omega_d (K+1)^2} \, .
\end{equation}

\section*{Density profiles at the plus-end}
Analogously to the minus-end, we now evaluate the density profiles and ensuing DW at the plus-end. Because the tip steadily depolymerizes, the calculations have to be performed in a comoving frame which is introduced first.

\subsection*{Comoving frame}

In the comoving frame, the above defined last lattice site $L$, i.e. the plus-end, is defined as the first site of the MT.
This is equivalent to reverting the motor movement. Since molecular motors in the comoving frame move towards the first site of the lattice, the transport current changes sign $J^{T}_i=-J^{T;Co}_i$
\begin{eqnarray}
J^{T;Co}_i:= \avg{n_{i+1} (1-n_i)} -  \avg{n_{i} (1-n_{i-1})} \, ,
\end{eqnarray}
using the mean-field approximation as introduced above this leads to 
\begin{eqnarray}
J^{T;Co}(x)=-(2\rho-1)\partial_x\rho(x) \,.
\end{eqnarray}
The particle adsorption/desorption current $J_i^{La}$ is unaffected. However, there is another contribution to the current balance in the comoving frame due to depolymerization:
Similar to the above definitions of the local currents, a local current which accounts for depolymerization in the comoving frame arises
\begin{equation}
J_i^{Co}=\delta(n_{i+1}-n_i)\, .
\end{equation}
Employing a mean-field approximation this expression simplifies to,
\begin{equation}
J^{Co}(x)=\delta\partial_x\rho(x)\,.
\end{equation}
This current term can be understood as follows. Due to the depolymerizing activity of a motor at the plus-end, in the comoving frame all motors on the MT simultaneously approach the plus-end.
In summary, by introducing a comoving frame the mean-field equation for the density at the MT plus-end $\rho_+(x)$ is obtained. In the steady state it reads
\begin{equation}
(2\rho_+ -1 -\delta)\partial_x\rho_+ +c \omega_a (1-\rho_+) - \omega_d\rho_+ = 0\,.
\label{eq:current_com}
\end{equation}

\subsection*{Density profiles}

The above equation is solved in close analogy to Eq.~\ref{eq:tasep_lk}. In terms of a rescaled density
 \begin{equation}
\sigma_+(x)=\frac{2\rho_+(x) - 2\frac{K}{K+1}}{2\frac{K}{K+1}-(1+\delta)}\,,
\end{equation}
the rescaled differential equation reads
\begin{equation}
\sigma^\prime_+(x)+\partial_x\ln|\sigma_+(x)|=\frac{\omega_d(K+1)^2}{K-1-(K+1)\delta}\, .
\end{equation}
The exact solutions to this equation are compared to stochastic simulations in the main text (Fig.~5{\it A}).

The solutions of this equation for $\delta=0$ are discussed in~\citep{PAR04S} and in parts above. However, in the case of depolymerization, i.e. for $\delta>0$, two special solutions exist:
Depending on the density of motors on the MT two classes of solutions for the density at the plus-end $\rho_+(x)$ can be distinguished. These are wedge-like or traffic jam density profiles; see Fig.~5{\it A} in the main text. The boundary condition for these qualitatively distinct density profiles is $\rho_+(L)=1$.
Defining
\begin{equation}
Y(x)=|\sigma_+(L)|\exp\left\{{\frac{\omega_d (K+1)^2}{K-1-\delta(K+1)}(x-L)+\sigma_+(L)}\right\},
\end{equation}
the two solutions for density profiles in the main text  (black lines in Fig.~5{\it A}) are 
\begin{eqnarray}
\rho(x)=
\begin{cases}
\rho_{La}+\frac{1}{2}(2\rho_{La}-(1+\delta)) W_{0}(Y(x))& \text{wedge-like}\\
\rho_{La}+\frac{1}{2}(2\rho_{La}-(1+\delta)) W_{-1}(-Y(x))& \text{traffic jam}.
\end{cases}
\end{eqnarray}
$W_0$ and $W_{-1}$ denote the first and the second real branch of the Lambert function.
The reason for the form of these two solutions is the bottleneck~\citep{PIE06S} arising due to depolymerization (see main text). This bottleneck fixes the value of the tip density to its maximum $\rho_+(L)=1$ for $\rho_{La}>\delta$. The transition from a traffic jam to a wedge-like density profile is thus not boundary-induced, i.e. due to a particular value of $\rho_+(L)<1$, but may be attributed to the depolymerizing activity of motors at the plus-end. As discussed in the main text, this transition is sharp and can be quantified in terms of $\rho_{La}$ and $\delta$; see Fig.~5{\it B}.

\subsection*{Domain wall position at the plus-end }

As already shown in the main text, microscopic jams can substantially influence the depolymerization dynamics. For large bulk concentrations, this perturbation no longer remains a local spike, but affects the profile on a macroscopic scale ~\citep{PIE06S}. Because the perturbation is macroscopic we can again use a hydrodynamic description, now with the boundary condition $\rho_+(L) = 1$.  Close to the plus-end Eq.~\ref{eq:current_com} gives an approximately linear profile
\begin{equation}
\rho_+ (x) \approx 1 - \frac{\omega_d}{1-\delta} (L-x) \, .
\label{eq:linear_tip_profile}
\end{equation}
The slope increases with increasing depolymerization rate $\delta$, c.f. Fig.~5{\it A} in the main text. In close analogy with the discussion for the minus-end there are two scenarios for concatenating this linear profile with the Langmuir density. Here, for large enough Langmuir density and/or small enough depolymerization rates, we obtain a wedge-like profile with a matching point given by $\rho_+ (d_w^+) = \rho_\text{La}$: 
\begin{equation}
d_w^+ \approx L - \frac{1-\delta}{\omega_d (1+K)} \, ;
\end{equation}
compare the green curves in Fig.~5{\it A} in the main text. Upon increasing the depolymerization rate or decreasing the Langmuir density a DW emerges whose position can be determined using current conservation, $\rho_+(d_+)  = 1-\rho_\text{La} + \delta$:
\begin{equation}
d^+ \approx L - \left[ \frac{K}{1+K}  - \delta \right] \frac{1-\delta}{\omega_d} \, ;
\end{equation}
compare the red curves in Fig.~5{\it A}. The DW is most pronounced for $K<1$. The height of the DW vanishes as $K$ approaches the threshold value 
\begin{equation}
K_c^+= \frac{1+\delta}{1-\delta} 
\end{equation}
from below. Equivalently, for a given $K$, the critical depolymerization rate reads
\begin{equation}
\delta_c = \frac{K-1}{K+1} = 2 \rho_\text{La} -1 
              = \frac{c \, \omega_a - \omega_d}{c \, \omega_a + \omega_d} \, .
\end{equation}
Together with the condition for spikes, $\rho_\text{La} < \delta$, this relation organizes the shapes of the density profiles into three classes: Microscopic jams at the tip, wedge profiles and DW profiles.

\section*{Depolymerization dynamics of the antenna profile}

In the main text we have discussed how a spatially uniform density $\rho_\text{La}$ affects depolymerization dynamics. Here we briefly show that our approach is also applicable to linear antenna profiles, i.e. for MTs shorter than a certain threshold length, $L<\ell_-$, cf. Eq.~\eqref{eq:length_threshold} and main text. Just as in the main text we equate the particle loss current due to depolymerization,
\begin{equation}
J_\text{depol}(x) =  \left[ 1-\rho_-(x) \right]  \rho_+(x) \delta \, ,
\label{eq:flux1}
\end{equation}
and the particle flux towards the plus-end,
\begin{equation}
J_{-}(x)=  \left[ 1-\rho_-(x) \right]  \rho_-(x) \, ,
\label{eq:flux2}
\end{equation}
and find
\begin{equation}
\rho_+(x)=\begin{cases} 
\frac{\rho_-(x)}{\delta} & \text{for} \; L<\frac{\delta}{c\omega_a} \, , \cr
1 & \text{for} \; L>\frac{\delta}{c\omega_a}\, .
\end{cases}
\label{eq:cases}
\end{equation}
Since, according to Eq.~(2) in the main text, the density at the plus-end determines the depolymerization speed, $V_\text{depol} = \delta \rho_+ (x)$, the  position-dependence of the tip density maps to a length-dependence of the polymerization speed. For MTs shorter than $\ell_-$ but longer than a certain depolymerization length $\ell_d=\delta/c\omega_a$ the depolymerization speed is length-independent
\begin{equation}
V_\text{depol}=\rho_+\delta =\delta \, .
\end{equation}
Analogously to the result  for constant bulk densities, this result shows that for MTs longer than the depolymerization length $\ell_d$ the dynamics of depolymerization of the antenna profile can not be distinguished from the dynamics as induced by a flat density profile.  In contrast, at a MT length shorter than $\ell_d$ the depolymerization speed becomes length-dependent and follows the shape of the antenna density profile $\rho_-(x)$:
\begin{equation}
V_\text{depol}(L)=\rho_+(L)\delta=\rho_-(L) \, .
\end{equation}
These results generalize the rate-limited and density-limited regimes discussed in the main text to non-uniform densities. Moreover, they show that once filaments become shorter than $\ell_-$, i.e. the density profile is antenna-like, there is a second spike-induced length scale $\ell_d$ which is the relevant length scale for the onset of length-dependent depolymerization of MTs.  

Combining these results with the analogous conditions for the Langmuir plateau discussed in the main text, leads to the depolymerization regimes summarized in Fig.~5C and Table~\ref{table:regimes}. Simply put, the depolymerization dynamics changes from rate-limited to density-limited when the \emph{bulk density} falls below the threshold density $\delta$: $\rho_-(L) \leq \delta$. For Langmuir densities below the threshold density, $\rho_\text{La} < \delta$, the bulk density remains below the threshold density for the whole MT length such that the depolymerization dynamics is always density-limited and given by: $V_\text{depol} = \rho_- (L)$.  This corresponds to regime ($\alpha$) in Fig.~5C. For Langmuir densities above the threshold, $\rho_\text{La} >\delta$, the depolymerization dynamics is rate-limited in the Langmuir plateau and given by $V_\text{depol} = \delta$. In the antenna-like regime of the density profile, i.e.\ for $L \leq \ell^-$, we have to distinguish between two cases: (i) $K>1$ ($\rho_\text{La}>0.5$) where the \emph{bulk} density profile exhibits a domain wall, and (ii) $K<1$ ($\rho_\text{La}<0.5$) where the \emph{bulk} density profile is wedge-like. In the latter case, the bulk-density profile changes slowly and hence remains above the threshold $\delta$ for some time even below $L=\ell^-$. Only for MTs shorter than $\ell_d$, given by $\rho_- (\ell_d) = \delta$, the dynamics changes from rate- to density-limited. This corresponds to regime ($\gamma$) in Fig.~5C.  In contrast, for $K>1$ ($\rho_\text{La} > 0.5$), the bulk density $\rho_- (x)$ exhibits a discontinuous jump from the Langmuir density to $1-\rho_\text{La}$ right at $\ell^-$. If this maximum value of the antenna-like profile is less than the threshold density, $1-\rho_\text{La} < \delta$, then the depolymerization dynamics discontinuously switches from rate-limited to density-limited. This defines regime ($\beta$) in Fig.~5C. Otherwise, if $1-\rho_\text{La} > \delta$, we are back to regime ($\gamma$). In summary, all regimes show a constant polymerization speed for long MTs in the Langmuir plateau. Depending on the relative magnitude of the Langmuir density and the depolymerization rate this regime is either density-limited and given by $\rho_\text{La}$ or rate-limited and given by $\delta$, cf. second column in Table~\ref{table:regimes}. In all scenarios the dynamics becomes length-dependent at some scale which is, however, different. While for regimes $\alpha$ and $\beta$, it coincides with the beginning of the antenna-like density profile $\ell^-$, it is given by $\ell_d$ for regime ($\gamma$) cf. third column in Table~\ref{table:regimes}.
\begin{table}[h]
\centering
\begin{tabular}{|c|c|c|c|c|}
\hline
Regime & Condition& Constant $V_\text{depol}$& Critical MT length \\ \hline\hline
$(\alpha)$ &  $\rho_\text{La}<\delta$&$V_\text{depol}=\rho_\text{La}$& $\ell_-$ \\ \hline
$(\beta)$ &  $\rho_\text{La}>\delta>1-\rho_\text{La}$&$V_\text{depol}=\delta$& $\ell_-$ \\ \hline
$(\gamma)$ & else &$V_\text{depol}=\delta$& $\ell_d$  \\ \hline
\hline
\end{tabular}
\caption{Summary of the similarities and differences in the depolymerization regimes $(\alpha),~(\beta)$ and $(\gamma)$. While for MTs longer than a critical length (third column) the depolymerization speed is constant (second column), it becomes length-dependent below this critical length and is then given by $V_\text{depol} = \rho_- (L)$. \label{table:regimes}}
\end{table}

\section*{Numerical implementation}
\label{sec:methods}

The stochastic dynamics of the individual-based model was simulated using a Gillespie algorithm~\citep{GIL76S} and employing the rates introduced above. Note that this method provides the mathematically exact stochastic dynamics. This is essential for the investigation of dynamic phenomena like length-dependent shortening.

In Fig.~2{\it A}-{\it D}., our simulations of the motor traffic started from an initial condition where the MT lattice was empty and subsequently filled up with motors triggering the depolymerization dynamics. To visualize time-dependent MT length and motor densities in one kymograph we implemented 14 protofilaments and averaged the motor intensities and MT lengths, see Fig.~2{\it A}-{\it C}. 
In detail, the visualization of kymographs was achieved as described in the following. From stochastic simulation data of the MT, each second the occupation numbers of motors along the MT $n_i^{(j)}$, where $j$ indexes the 14 protofilaments of the MT, were evaluated and converted to $RGB$ color values:
\begin{eqnarray}
R_i=\tfrac{\sum_{j=1}^{14} 1-n_i^{(j)}}{14} ,\;\;\; G_i=\tfrac{\sum_{j=1}^{14} n_i^{(j)}}{14} ,\;\;\;B_i=0.
\end{eqnarray}
These values display the density of motors as green and the uncovered MT surface as red. Hence, if the MT is completely empty it is red, while at complete motor coverage it is green.

Steady state motor densities as shown in Figs.~2{\it E} and 5{\it A} were obtained by time-averaging over $10^4$ independent realizations after an equilibration time of 2000 time steps; note that for a constant lattice size time and ensemble averages yield identical results~\citep{PAR03S}. In Fig.~5{\it A}, the density profiles were recorded in the comoving frame of the MT plus-end, while density profiles in Fig.~2{\it E} were recorded in the rest frame of the MT minus-end. In  Fig.~2{\it E}, the density profiles resulting from the minus end without any influences from the plus end are shown. This can be viewed as an infinitely long lattice. To simulate such a lattice, we chose the following boundary conditions. We neglected depolymerization as it arises at infinity. Further, we set the exiting rate for motors at the last site equal to $1-\rho_\text{La}$. Then, the transport behavior at the tip is the same as on the lattice, if the Langmuir density is reached, $\rho(x)=\rho_\text{La}$.

In the second part of the article we focus on the dependence of the depolymerization speed on the motor density. To this end, simulations were performed in a comoving frame where the density at the minus-end was fixed to the Langmuir density. This was achieved by extending the lattice one site to left with each depolymerization step and filling the thereby created site with the probability $\rho_\text{La}$. This procedure may also be interpreted as an infinite MT allowing to observe motor dynamics at the MT tip without perturbations arising from the length-dependent depolymerization regime. 

We measured the mean end-residence time of individual motors at the plus-end $\tau_\mathrm{res}$ and the mean lifetime of the terminal tubulin dimer~$\tau_d$. Data of these were obtained by averaging over $10^4$ time steps $\tau$, cf. Figs.~4 and~6. 

\section*{\bf How kymographs become independent of the depolymerization rate}

\begin{figure} [h]
\centering
\includegraphics*[width=3.25 in]{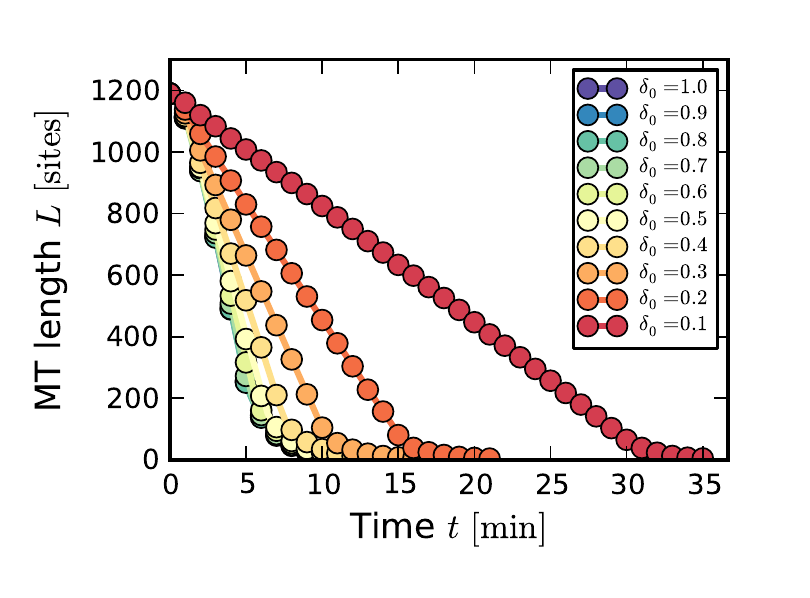}
\caption{\emph{Kymographs as they become independent of the depolymerization rate~$\delta$.} Time-space plots of depolymerizing MTs for different depolymerization rates are shown, ranging from $\delta=0.1\,\nu$ (red) to $1.0\,\nu$ (blue), the latter value corresponds to the motor speed of $v=6.35\,\mathrm{sites}\,\mathrm{s}^{-1}$.
For slow depolymerization rate, $\delta \lessapprox 0.5$, the depolymerization speed is related to the microscopic depolymerization rate $\delta$, whereas for rates  $\delta \gtrapprox 0.5$ the depolymerization speed is independent of the depolymerization rate but depends on the density of motors on the MT, as outlined in the main text.
}
\label{fig:kymo}
\end{figure}

In Fig.~\ref{fig:kymo} we provide data that explicitly shows the parameter independence of MT depolymerization. This results has been generalized in the main text to all possible motor concentrations and depolymerization rates.
\setcounter{Sfigure}{1}

\section*{\bf Comparison with experiments: dependence of the depolymerization speed on the bulk flux and bulk density}

\begin{figure} [t]
\centering
\includegraphics*[width=3.25 in]{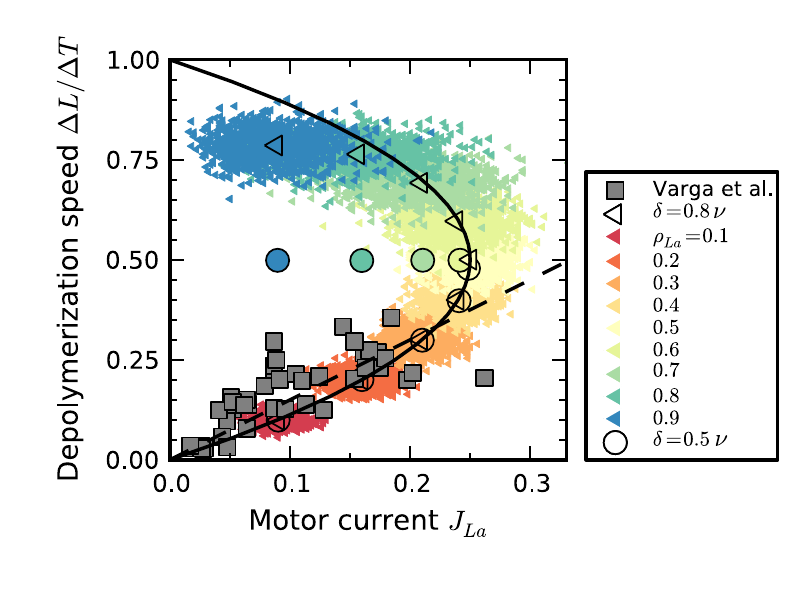}
\caption{\emph{Depolymerization speed $\Delta L/\Delta T$ as a function of the bulk motor current on the MT, $J_\text{La}$.} Data from stochastic simulations for two different depolymerization rates $\delta=0.5\,\nu$ ($\circ$) and $0.8\,\nu$ ($\triangleleft$) are shown, each for a set of Langmuir densities (colors) corresponding to concentrations $c=0.16\ldots 13\,\text{nM}$. To illustrate the effect of statistics, small symbols show individual measurements as obtained from $10^3$ fixed time measurements $\Delta T=500\tau$, while large symbols indicate their mean $\avg{\Delta L/\Delta T}$. Good agreement with experimental data ($\Box$) measured in the low density regime is found~\citep{VAR09S}.
Here, the theoretical prediction given by $J_\text{La}$ (solid) is hardly discernible from a linear best fit to experiments (dashed).
}
\label{fig:depolimerization_speed_current}
\end{figure}

Experimentally it was found that the depolymerization speed is linearly correlated with the flux of molecular motors towards the plus-end~\citep{VAR09S}. We have collected data from our simulations similar to experiments. Figure~\ref{fig:depolimerization_speed_current} shows a scatter plot for the depolymerization speed as a function of the bulk flux of motors, $J_\text{La}$, for two values of the microscopic depolymerization rate $\delta$.
\setcounter{Sfigure}{2}

The noise in the ensemble of realizations has two sources. The bulk current fluctuates since the Langmuir kinetics responsible for the bulk density $\rho_\text{La}$ is a stochastic process. The depolymerization speeds vary from realization to realization because the depolymerization kinetics is a Poisson-like process. Also shown in Fig.~\ref{fig:depolimerization_speed_current} are ensemble averages. These mean values, as predicted in the main text, show the following behavior. For a macroscopic depolymerization speed lower than the depolymerization rate, ${\Delta L}/{\Delta T} < \delta a$, it is density-limited and identical to the bulk density:
\begin{equation}
\frac{\Delta L}{\Delta T} = \rho_\text{La} v \quad \text{for} \quad \rho_\text{La} < \delta \tau\, .
\end{equation}
Rewriting this relation in terms of the bulk current means that the data should fall on the parabola displayed as the solid curve in Fig.~\ref{fig:depolimerization_speed_current}. For low densities, $\rho_\text{La} \lessapprox 0.25$, where crowding effects are weak, this implies ${\Delta L}/{\Delta T} = \rho_\text{La} v \approx J_\text{La}$ as observed experimentally \citep{VAR09S}; see Fig.~\ref{fig:depolimerization_speed_current}. 

As the bulk density is increased two things happen. First, crowding effects become important invalidating the linear relationship between bulk current and depolymerization speed. It would be interesting to test our prediction that the depolymerization speed is linear in the bulk density by using higher motor concentrations or changed biochemical rates such that $K$ becomes significantly larger than $1$.

Second, if $\rho_\text{La} > \delta \tau$, the depolymerization speed becomes rate-limited:
\begin{equation}
\frac{\Delta L}{\Delta T} = a \delta \quad \text{for} \quad \rho_\text{La} > \delta \tau \, .
\end{equation}
This puts an obvious upper bound on the depolymerization speed. It cannot become larger than the microscopic rate of depolymerization at the plus-end. If the depolymerization rate is larger than the hopping rate of the molecular motors, $\delta > \nu$, the depolymerization speed is, for all possible values of the bulk density, strictly given by the bulk density.


\begin{thebibliography}{44}
\providecommand{\natexlab}[1]{#1}

\bibitem{HAY01}
Hayles, J., and P.~Nurse. 2001.
\newblock A journey into space.
\newblock \emph{Nat. Rev. Mol. Cell Biol.} 2:647--656.

\bibitem{TOL10}
Toli\'c-N{\o}rrelykke, I.~M. 2010.
\newblock Force and length regulation in the microtubule cytoskeleton: lessons
  from fission yeast.
\newblock \emph{Curr. Opin. Cell Biol.} 22:21--28.

\bibitem{SHA00}
Sharp, D., G.~Rogers, and J.~Scholey. 2000.
\newblock Microtubule motors in mitosis.
\newblock \emph{Nature}. 407:41--47.

\bibitem{KAR01}
Karsenti, E., and I.~Vernos. 2001.
\newblock Cell cycle - the mitotic spindle: A self-made machine.
\newblock \emph{Science}. 294:543--547.

\bibitem{EGG06}
Eggert, U.~S., T.~J. Mitchison, and C.~M. Field. 2006.
\newblock Animal cytokinesis: From parts list to mechanisms.
\newblock \emph{Annu. Rev. Biochem.} 75:543--566.

\bibitem{HIR09}
Hirokawa, N., Y.~Noda, Y.~Tanaka, and S.~Niwa. 2009.
\newblock Cytoskeletal motors: Kinesin superfamily motor proteins and
  intracellular transport.
\newblock \emph{Nat. Rev. Mol. Cell Biol.} 10:682--696.

\bibitem{MIT84}
Mitchison, T., and M.~Kirschner. 1984.
\newblock Dynamic instability of microtubule growth.
\newblock \emph{Nature}. 312:237--242.

\bibitem{DOG93}
Dogterom, M., and S.~Leibler. 1993.
\newblock Physical aspects of the growth and regulation of microtubule
  structures.
\newblock \emph{Phys. Rev. Lett.} 70:1347--1350.

\bibitem{DES97}
Desai, A., and T.~Mitchison. 1997.
\newblock Microtubule polymerization dynamics.
\newblock \emph{Annu. Rev. Cell Dev. Biol.} 13:83--117.

\bibitem{HOW03}
Howard, J., and A.~Hyman. 2003.
\newblock Dynamics and mechanics of the microtubule plus end.
\newblock \emph{Nature}. 422:753--758.

\bibitem{WOR05}
Wordeman, L. 2005.
\newblock Microtubule-depolymerizing kinesins.
\newblock \emph{Curr. Opin. Cell Biol.} 17:82--88.

\bibitem{HOW07}
Ho\-ward, J., and A.~A. Hyman. 2007.
\newblock Microtubule polymerases and depolymerases.
\newblock \emph{Curr. Opin. Cell Biol.} 19:31--35.

\bibitem{HOW09}
Howard, J., and A.~A. Hyman. 2009.
\newblock Growth, fluctuation and switching at microtubule plus ends.
\newblock \emph{Nat. Rev. Mol. Cell Biol.} 10:569--574.

\bibitem{HEL06}
Helenius, J., G.~Brouhard, Y.~Kalaidzidis, S.~Diez, and J.~Howard. 2006.
\newblock The depolymerizing kinesin {MCAK} uses lattice diffusion to rapidly
  target microtubule ends.
\newblock \emph{Nature}. 441:115--119.

\bibitem{VAR06}
Varga, V., J.~Helenius, K.~Tanaka, A.~A. Hyman, T.~U. Tanaka, and J.~Howard.
  2006.
\newblock Yeast kinesin-8 depolymerizes microtubules in a length-dependent
  manner.
\newblock \emph{Nat. Cell Biol.} 8:957--962.

\bibitem{GUP06}
Gupta, M.~L., P.~Carvalho, D.~M. Roof, and D.~Pellman. 2006.
\newblock Plus end-specific depolymerase activity of {K}ip3, a kinesin-8
  protein, explains its role in positioning the yeast mitotic spindle.
\newblock \emph{Nat. Cell Biol.} 8:913--923.

\bibitem{MAY07}
Mayr, M.~I., S.~H\"ummer, J.~Bormann, T.~Gr\"uner, S.~Adio, G.~Woehlke, and
  T.~U. Mayer. 2007.
\newblock The human kinesin {K}if18a is a motile microtubule depolymerase
  essential for chromosome congression.
\newblock \emph{Curr. Biol.} 17:488--498.

\bibitem{STU08}
Stumpff, J., G.~V. Dassow, M.~Wagenbach, C.~Asbury, and L.~Wordeman. 2008.
\newblock The kinesin-8 motor {K}if18a suppresses kinetochore movements to
  control mitotic chromosome alignment.
\newblock \emph{Dev. Cell}. 14:252--262.

\bibitem{DU10}
Du, Y., C.~A. English, and R.~Ohi. 2010.
\newblock The kinesin-8 {K}if18a dampens microtubule plus-end dynamics.
\newblock \emph{Curr. Biol.} 20:374--380.

\bibitem{UNS08}
Unsworth, A., H.~Masuda, S.~Dhut, and T.~Toda. 2008.
\newblock Fission yeast kinesin-8 {K}lp5 and {K}lp6 are interdependent for
  mitotic nuclear retention and required for proper microtubule dynamics.
\newblock \emph{Mol. Biol. Cell}. 19:5104--5115.

\bibitem{TIS09}
Tischer, C., D.~Brunner, and M.~Dogterom. 2009.
\newblock Force-and kinesin-8-dependent effects in the spatial regulation of
  fission yeast microtubule dynamics.
\newblock \emph{Mol. Syst. Biol.} 5:250.

\bibitem{GRI09}
Grissom, P., T.~Fiedler, E.~Grishchuk, D.~Nicastro, R.~West, and J.~R.
  McIntosh. 2009.
\newblock Kinesin-8 from fission yeast: A heterodimeric, plus-end-directed
  motor that can couple microtubule depolymerization to cargo movement.
\newblock \emph{Mol. Biol. Cell}. 20:963.

\bibitem{VAR09}
Varga, V., C.~Leduc, V.~Bormuth, S.~Diez, and J.~Howard. 2009.
\newblock Kinesin-8 motors act cooperatively to mediate length-dependent
  microtubule depolymerization.
\newblock \emph{Cell}. 138:1174--1183.

\bibitem{GAR08}
Gardner, M.~K., D.~C. Bouck, L.~V. Paliulis, J.~B. Meehl, E.~T. O'Toole,
  J.~Haase, A.~Soubry, A.~P. Joglekar, M.~Winey, E.~D. Salmon, K.~Bloom, and
  D.~J. Odde. 2008.
\newblock Chromosome congression by kinesin-5 motor-mediated disassembly of
  longer kinetochore microtubules.
\newblock \emph{Cell}. 135:894--906.

\bibitem{FOE09}
Foethke, D., T.~Makushok, D.~Brunner, and F.~N\'ed\'elec. 2009.
\newblock Force- and length-dependent catastrophe activities explain interphase
  microtubule organization in fission yeast.
\newblock \emph{Mol. Syst. Biol.} 5:241.

\bibitem{KLU08}
Klumpp, S., Y.~Chai, and R.~Lipowsky. 2008.
\newblock Effects of the chemomechanical stepping cycle on the traffic of
  molecular motors.
\newblock \emph{Phys. Rev. E}. 78:041909.

\bibitem{HOW96}
Howard, J. 1996.
\newblock The movement of kinesin along microtubules.
\newblock \emph{Annu. Rev. Physiol.} 58:703--729.

\bibitem{RAY93}
Ray, S., E.~Meyh{\"o}fer, R.~Milligan, and J.~Howard. 1993.
\newblock Kinesin follows the microtubule protofilament axis.
\newblock \emph{J. Cell Biol.} 121:1083--1093.

\bibitem{PAR03}
Parmeggiani, A., T.~Franosch, and E.~Frey. 2003.
\newblock Phase coexistence in driven one-dimensional transport.
\newblock \emph{Phys. Rev. Lett.} 90:086601.

\bibitem{PAR04}
Parmeggiani, A., T.~Franosch, and E.~Frey. 2004.
\newblock Totally asymmetric simple exclusion process with langmuir kinetics.
\newblock \emph{Phys. Rev. E}. 70:046101.

\bibitem{LIP01}
Lipowsky, R., S.~Klumpp, and T.~Nieuwenhuizen. 2001.
\newblock Random walks of cytoskeletal motors in open and closed compartments.
\newblock \emph{Phys. Rev. Lett.} 87:108101.

\bibitem{KLU03}
Klumpp, S., and R.~Lipowsky. 2003.
\newblock Traffic of molecular motors through tube-like compartments.
\newblock \emph{J. Stat. Phys.} 113:233--268.

\bibitem{PIE06}
Pierobon, P., M.~Mobilia, R.~Kouyos, and E.~Frey. 2006.
\newblock Bottleneck-induced transitions in a minimal model for intracellular
  transport.
\newblock \emph{Phys. Rev. E}. 74:031906.

\bibitem{TEL09}
Telley, I.~A., P.~Bieling, and T.~Surrey. 2009.
\newblock Obstacles on the microtubule reduce the processivity of {K}inesin-1
  in a minimal in vitro system and in cell extract.
\newblock \emph{Biophys. J.} 96:3341--3353.

\bibitem{GOV08}
Govindan, B.~S., M.~Gopalakrishnan, and D.~Chowdhury. 2008.
\newblock Length control of microtubules by depolymerizing motor proteins.
\newblock \emph{Europhys. Lett.} 83:40006.

\bibitem{BRU09}
Brun, L., B.~Rupp, J.~J. Ward, and F.~Nedelec. 2009.
\newblock A theory of microtubule catastrophes and their regulation.
\newblock \emph{Proc. Natl. Acad. Sci. USA}. 106:21173--21178.

\bibitem{HOU09}
Hough, L.~E., A.~Schwabe, M.~A. Glaser, J.~R. McIntosh, and M.~D. Betterton.
  2009.
\newblock Microtubule depolymerization by the kinesin-8 motor {K}ip3p: A
  mathematical model.
\newblock \emph{Biophys. J.} 96:3050--3064.

\bibitem{KLE05}
Klein, G.~A., K.~Kruse, G.~Cuniberti, and F.~J\"ulicher. 2005.
\newblock Filament depolymerization by motor molecules.
\newblock \emph{Phys. Rev. Lett.} 94:108102.

\bibitem{VIL01}
Vilfan, A., E.~Frey, F.~Schwabl, M.~Thormahlen, Y.~Song, and E.~Mandelkow.
  2001.
\newblock Dynamics and cooperativity of microtubule decoration by the motor
  protein kinesin.
\newblock \emph{J. Mol. Biol.} 312:1011--1026.

\bibitem{FRE02}
Frey, E., and A.~Vilfan. 2002.
\newblock Anomalous relaxation kinetics of biological lattice-ligand binding
  models.
\newblock \emph{Chem. Phys.} 284:287--310.

\bibitem{FRE04}
Frey, E., A.~Parmeggiani, and T.~Franosch. 2004.
\newblock Collective phenomena in intracellular processes.
\newblock \emph{Genome Informatics}. 15(1):46--55.

\bibitem{COO10}
Cooper, J.~R., M.~Wagenbach, C.~L. Asbury, and L.~Wordeman. 2010.
\newblock Catalysis of the microtubule on-rate is the major parameter
  regulating the depolymerase activity of {MCAK}.
\newblock \emph{Nat. Struct. Mol. Biol.} 17:77--82.

\bibitem{KIN01}
Kinoshita, K., I.~Arnal, A.~Desai, D.~Drechsel, and A.~Hyman. 2001.
\newblock Reconstitution of physiological microtubule dynamics using purified
  components.
\newblock \emph{Science}. 294:1340--1343.

\bibitem{BRO08}
Brouhard, G.~J., J.~H. Stear, T.~L. Noetzel, J.~Al-Bassam, K.~Kinoshita, S.~C.
  Harrison, J.~Howard, and A.~A. Hyman. 2008.
\newblock {XMAP}215 is a processive microtubule polymerase.
\newblock \emph{Cell}. 132:79--88.

\end{thebibliography}

\begin{thebibliography}{7}
\providecommand{\natexlab}[1]{#1}

\bibitem{VAR09S}
Varga, V., C.~Leduc, V.~Bormuth, S.~Diez, and J.~Howard. 2009.
\newblock Kinesin-8 motors act cooperatively to mediate length-dependent
  microtubule depolymerization.
\newblock \emph{Cell}. 138:1174--1183.

\bibitem{PAR03S}
Parmeggiani, A., T.~Franosch, and E.~Frey. 2003.
\newblock Phase coexistence in driven one-dimensional transport.
\newblock \emph{Phys. Rev. Lett.} 90:086601.

\bibitem{PAR04S}
Parmeggiani, A., T.~Franosch, and E.~Frey. 2004.
\newblock Totally asymmetric simple exclusion process with langmuir kinetics.
\newblock \emph{Phys. Rev. E}. 70:046101.

\bibitem{COR96S}
Corless, R., G.~Gonnet, D.~Hare, D.~Jeffrey, and D.~Knuth. 1996.
\newblock On the lambert $w$ function.
\newblock \emph{Adv. Comput. Math.} 5:329--359.

\bibitem{PIE06S}
Pierobon, P., M.~Mobilia, R.~Kouyos, and E.~Frey. 2006.
\newblock Bottleneck-induced transitions in a minimal model for intracellular
  transport.
\newblock \emph{Phys. Rev. E}. 74:031906.

\bibitem{VAR06S}
Varga, V., J.~Helenius, K.~Tanaka, A.~A. Hyman, T.~U. Tanaka, and J.~Howard.
  2006.
\newblock Yeast kinesin-8 depolymerizes microtubules in a length-dependent
  manner.
\newblock \emph{Nat. Cell Biol.} 8:957--962.

\bibitem{GIL76S}
Gillespie, D.~T. 1976.
\newblock Stochastic simulations of chemical processes.
\newblock \emph{J. Comp. Phys.} 22:403--434.

\end{thebibliography}
\end{document}